%% file: Adecay.tex
\documentclass{paper}
\usepackage{fmslash}
\usepackage{feynarts}
\usepackage{feynrules}
\def\FALabelSize{\footnotesize}
\def\FADiagramLabelSize{\footnotesize}

\newcommand{\Lcal}{\mathcal{L}}

\newcommand{\Dslash}{\fmslash[1mu]{D}}

\newcommand{\cc}{\mathsf{c}}

\renewcommand{\feynrulesvertexiii}[4]{%
  \noindent $#1$ $#3$ $#2$:\nopagebreak\\[\medskipamount]
  \hspace*{2em}$\displaystyle #4$\par\bigskip%
}

\begin{document}
\begin{titlepage}
  \begin{flushright}
    PITHA 09/19
  \end{flushright}        
\vspace{0.01cm}
% \vspace{2cm}
\begin{center}
{\Large{\bf Pseudoscalar Higgs Boson Decays \\ \vspace{0.5cm} into W and Z Bosons Revisited}}
%\author{W.\ Bernreuther\and P.\ Gonzalez\and M.\ Wiebusch}
 \par\vspace{1.5cm}
{\large{\bf Werner Bernreuther\footnote{Email:
{\tt breuther@physik.rwth-aachen.de}},
 Patrick Gonz\'alez\footnote{Email:
{\tt gonzalez@physik.rwth-aachen.de}},
 Martin Wiebusch\footnote{Email: {\tt mwiebusch@physik.rwth-aachen.de}} 
}}
\par\vspace{1cm}
Institut f\"ur Theoretische Physik, RWTH Aachen University, 52056 Aachen,
Germany\\
\par\vspace{3cm}
{\bf Abstract}\\
\parbox[t]{\textwidth}
{We examine, in a number of Standard Model extensions,
  whether the decays $A \to WW/ZZ$   of a neutral
pseudoscalar (Higgs) resonance can have branching ratios at the percent
level and we determine 
the possible size of $B(A \to WW/ZZ)$ relative to the respective branching
ratios of a scalar boson $H$. The branching ratios of the these decay
modes  and the total widths $\Gamma_A$, $\Gamma_H$ 
are computed in 
 the minimal supersymmetric extension of the SM,
 in  a type-II two-Higgs doublet extension (2HDM), in
 a 2HDM with 4 chiral fermion generations, in 
 a 2HDM with additional  heavy vector-like quarks,  and in  
  a top-color assisted technicolor model. We find that in
  the above non-supersymmetric models $B(A \to WW)$ can be about
  $2\%$, while  $B(A \to ZZ) \lesssim 10^{-3}.$ The ratio
  $B(A \to WW)/B(H \to WW)$ can be of order $10\%$ in large regions
  of the parameter space of the models.
 }
\end{center}
\vspace*{2cm}
\noindent
PACS number(s):  12.60.-i, 12.60.Fr,  12.60Jv, 12.60.Nz, 14.80.Cp  
\end{titlepage}
%
%
%
%%%%%%%%%%%%%%%%%%%%%%%%%%%%%%%%%%%%%%%%%%%%%%%%%%%%%%%%%%%%%%%%%%%%%%%%%%%%%%%%
\section{Introduction}
%%%%%%%%%%%%%%%%%%%%%%%%%%%%%%%%%%%%%%%%%%%%%%%%%%%%%%%%%%%%%%%%%%%%%%%%%%%%%%%%
%
%
The search for Higgs bosons or, more general, (spin-zero) resonances is among
the major physics goals of present-day collider physics, as the existence of
such resonances and exploration of their properties (quantum numbers, production
and decay modes) would yield decisive clues for unraveling the mechanism of
electroweak gauge symmetry breaking (EWSB).  There is an exhaustive
phenomenology of the production and decay modes of
the standard model (SM) Higgs boson; likewise, there are extensive theoretical
studies of these issues for spin-zero (Higgs) particles predicted by popular SM
extensions.  (For reviews see, e.g., \cite{Djouadi:2005gi} and
\cite{Djouadi05,Accomando:2006ga,Hill:2002ap}, respectively.)

For the SM Higgs boson $H$ with a mass $m_H \gtrsim 130$ GeV, signatures from
the decay modes\footnote{For state-of-the-art predictions for $H\to W W/Z Z \to
  4 \, {\rm fermions}$, see \cite{Bredenstein:2006ha}.}  $H\to W W^{(*)}/ Z
Z^{(*)}$ have the highest discovery potential for this particle at the Large
Hadron Collider (LHC) \cite{Asai:2004ws,Abdullin:2005yn}. Concerning
non-standard neutral Higgs particles the situation is, especially in view of
unknown model parameters, less clear as to which decay channel is, for a
specific production mode, the most promising one. However, as is well-known, for
a pseudoscalar Higgs boson $A$ it is expected that the decays $A\to W W/ Z Z$
are strongly suppressed. This is because the couplings $AVV$ $(V=W,Z)$ must be
loop-induced, and they turn out to be very small in two-Higgs doublet extensions
(2HDM) and the minimal supersymmetric extension (MSSM) of the SM
\cite{Mendez:1991gp,Gunion:1991cw}. In view of this conventional wisdom one
might be inclined to conclude that the discovery of a spin-zero resonance in the
$WW$ and/or $ZZ$ channel immediately suggests that this resonance is a scalar,
i.e., a $J^{PC}=0^{++}$ state. We hasten to add that many suggestions and
phenomenological studies have been made how the spin and the $CP$ parity of a
resonance can actually be measured in these channels \cite{Nelson:1986ki,
  Soni:1993jc, Skjold:1993jd, Barger:1993wt, Arens:1994wd, Choi:2002jk,
  Buszello:2002uu, Godbole:2007cn, Accomando:2006ga}.

In this paper, we examine  whether the decays of a neutral
pseudoscalar into $WW/ZZ$ can have branching ratios at the percent
level and analyse
the possible size of $B(A \to WW/ZZ)$ relative to the respective branching
ratios of a scalar boson $H$. We analyze this in
the context of several phenomenologically viable SM extensions.  Concerning the
scenario that EWSB is triggered by elementary Higgs fields, we choose a Higgs
sector which is composed of two ${\rm SU(2)}$ doublets. For the convenience of
the reader we first reexamine $B(A\to VV)$ in 2HDM and the MSSM. Our analysis
for models with additional heavy (exotic) quarks and leptons is new, to the best
of our knowledge. We choose top-color assisted technicolor models (TC2)
\cite{Hill:1994hp} as a paradigm for models that predict a relatively light
composite pseudoscalar.

We assume here that the dynamics of the EWSB sector of the Yukawa interactions
is such that the electrically neutral Higgs resonances are CP eigenstates in the
mass basis, at least to very good approximation. If the Higgs sector violates CP
then, as is well known, the neutral spin-zero mass eigenstates are, in general,
a mixture of a CP-odd and CP-even component, of which the latter has tree-level
couplings to $WW/ZZ$.  

The paper is organized as follows. In Sect.~\ref{sec:modres}, we compute for the above models
the branching ratios $B(A\to VV)$ for masses  $m_A \geq 200\,{\rm
  GeV}$. We estimate the maximal size of these branching ratios and the 
 total width $\Gamma_A$ within a given
model. For comparison we also determine  the branching ratios $H\to
VV$ of a scalar $H$ with mass $m_H \simeq m_A$ and its width $\Gamma_H$.  
 We conclude in Sect.~\ref{sec:conclusions}. In an appendix we list
for the convenience of the reader, for our model with vector-like quarks, the
couplings of these quarks to Higgs bosons.
%
%
%
%%%%%%%%%%%%%%%%%%%%%%%%%%%%%%%%%%%%%%%%%%%%%%%%%%%%%%%%%%%%%%%%%%%%%%%%%%%%%%%%
\section{Models and results} \label{sec:modres}
%%%%%%%%%%%%%%%%%%%%%%%%%%%%%%%%%%%%%%%%%%%%%%%%%%%%%%%%%%%%%%%%%%%%%%%%%%%%%%%%
%
%
First, we collect a few basic formulae needed in the next section.  Lorentz
covariance dictates that the amplitude for the decay of a spin-zero state $\phi$
into a pair of vector bosons, $\phi(p) \to V(p_1) V(p_2)$ ($V=W,Z$) contains in
general three form factors:
\begin{equation} \label{phivv}
    i {\cal M}_{VV}
  = f_{1,VV} \, \epsilon_1^*\cdot\epsilon_2^* + f_{2,VV}\,
    (p_1\cdot\epsilon_2^*)(p_2\cdot\epsilon_1^*)
   + f_{3,VV}\,\epsilon_{\mu\nu\rho\sigma}
   p_1^\mu  p_2^\nu\epsilon_1^{*\rho}\epsilon_2^{*\sigma}
   \eqpunct,
\end{equation}
where $\epsilon_1$ and $\epsilon_2$ are the polarization vectors of the vector
bosons. For a pseudoscalar, $\phi =A$, the form factors $f_{1,VV}$ and
$f_{2,VV}$ vanish if CP is conserved. Defining for this case the effective
dimensionless coupling $C_{VV}$ by
\begin{equation} \label{def:cvv}
  f_{3,VV}\equiv\frac{e}{s_W m_W}C_{VV}
  \eqpunct, 
 \end{equation}
where $s_W \, (c_W)$ denotes the sine (cosine) of the weak mixing angle, one
obtains for the decay rate of a pseudoscalar into a pair of on-shell vector
bosons:
\begin{equation}\label{eq:gen:Awidth}
    \Gamma(A\to VV)  
  = r_V \frac{ G_F}{8\sqrt{2} \pi} m_A^3 \beta_V^3  \, |C_{VV}|^2
  \eqpunct,
\end{equation}
where $\beta_V =(1 - 4 {m_V^2}/{m_A^2})^{\frac{1}{2}}$ and $r_V = 2 \, (1)$ for
$V=W \, (Z)$.

In gauge theories, or more general in renormalizable theories, the coupling
$f_{3,VV}$ is not present at tree level, but must be induced by loop
corrections.  
 In the bosonic sectors of the SM extensions considered in the next
section, C and P are separately conserved. Thus, the bosonic sectors of these
theories cannot induce, at any order, a non-zero $AVV$ coupling -- it requires
fermions which, through their C- and P-violating couplings to $W$ and $Z$
bosons, generate $f_{3,VV} \neq 0$ at 1-loop order. If the Yukawa couplings of
$A$ are of the Higgs-boson type, then $C_{VV}$ exhibits the well-known
non-decoupling behaviour of Higgs boson interactions. In this case, $C_{VV}$
does not vanish for heavy fermion(s) circulating in the loop, but becomes
independent of the internal mass(es).

For completeness we also list the rates of the decays of a scalar
(Higgs) boson $H$ into a $WW$ and $ZZ$, as we will compare the $A,H \to VV$
branching ratios in the next section.  The $HVV$ amplitude at Born level can be
parameterized by $ 2m_V^2 \sqrt{\sqrt{2}G_F}
h_{VV}\epsilon_1^*\cdot\epsilon_2^*$, where $h_{VV}$ is a dimensionless coupling
which is equal to one in the SM. The tree-level decay rate is
\begin{equation}\label{eq:gen:Hwidth}
  \Gamma(H\to VV)
  = r_V \frac{G_F}{16 \sqrt{2} \pi} m_H^3 \beta_V \left[
    \beta_V^2 +12\frac{m_V^4}{m_H^4}\right] h_{VV}^2
  \eqpunct.
\end{equation}

Below we shall discuss SM extensions with two ${\rm SU(2)}$ Higgs doublets as a
paradigm for an extended Higgs sector, with a dynamics such that the physical
spin-zero mass eigenstates are also CP eigenstates.  That is, the Higgs boson
spectrum contains, apart from a charged Higgs-boson pair $H^\pm$, two CP-even
states $h, H$ and a CP-odd state $A$. The mass of $A$ is assumed to be
 $m_A \geq 200 \, {\rm GeV}.$ For comparison, we
consider also the decay of the heavier of the two CP-even Higgs bosons, denoted
by $H$, into $W^+W^-$ and $ZZ$. 
We define the ratios
\begin{equation} \label{rvratio}
  R_V = \frac{B(A\to VV)}{B(H\to VV)}
  \eqpunct,
\end{equation}
for  comparing the respective branching ratios 
of  $A$ and $H$. 

We will also consider top-color assisted technicolor, which is a
phenomenologically viable model for ``dynamical'' gauge symmetry breaking. The
spin-zero particle spectra of this class of models contain a triplet of ``top
pions'', $\Pi^{0,\pm}$, of which $\Pi^0$ takes the role of the pseudoscalar $A$,
and a scalar bound state, the ``top Higgs'' $H_t$, with a mass of the order of a
few hundred GeV.

Throughout this paper we will  use the following SM parameters:
\begin{gather}
  1/\alpha_{\text{em}} = 137.036
  \quad,\quad
  \alpha_s = 0.118
  \eqpunct,\nonumber\\
  m_Z = \unit{91.19}{GeV}
  \quad,\quad
  m_W = \unit{80.40}{GeV}
  \eqpunct,\nonumber\\
  m_t = \unit{172.6}{GeV}
  \quad,\quad
  m_b = \unit{4.79}{GeV}
  \quad,\quad
  m_\tau = \unit{1.78}{GeV}
  \quad,\quad
  V_{tb} = 1
  \eqpunct.\label{eq:sm_parameters}
\end{gather}
In view of the small Yukawa couplings of the quarks and leptons of the
first and second generation, their interactions with the Higgs
resonances can be neglected in the analysis below.  For the
calculations we used  \texttt{FeynArts}
 \cite{Hahn:2000kx,Hahn:2001rv} in combination with 
 \texttt{FormCalc} and \texttt{LoopTools} 
    \cite{Hahn:1998yk,Hahn:2006qw}.
%
%
%%%%%%%%%%%%%%%%%%%%%%%%%%%%%%%%%%%%%%%%%%%%%%%%%%%%%%%%%%%%%%%%%%%%%%%%%%%%%%%%
\subsection{Two-Higgs doublet extensions}\label{sec:res:THDM}
%%%%%%%%%%%%%%%%%%%%%%%%%%%%%%%%%%%%%%%%%%%%%%%%%%%%%%%%%%%%%%%%%%%%%%%%%%%%%%%%
%
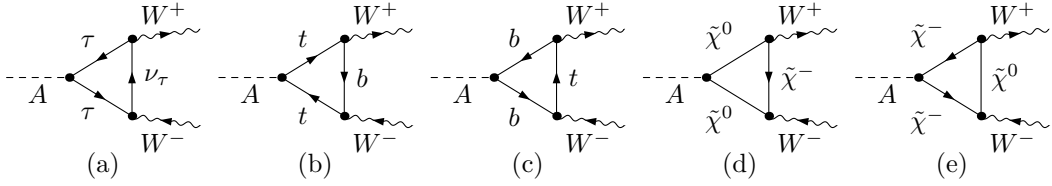
\begin{figure}
  \centering
  \input{dgmWW.tex}
  \caption{Diagrams that contribute in the MSSM to the decay $A\to WW$.  In the
    2HDM the 1-loop decay amplitude is due to the diagrams (a--c) only. As we
    consider decays to on-shell gauge bosons, diagrams involving the $AZ$
    transition amplitude are absent.}
  \label{fig:dgmWW}
\end{figure}
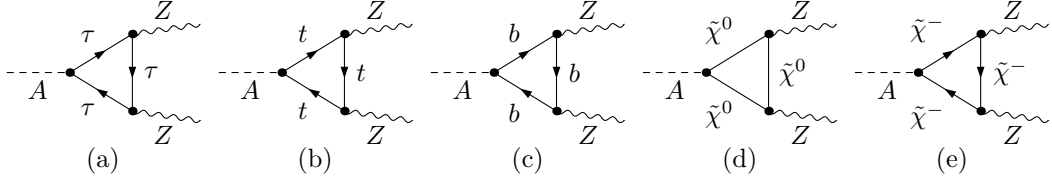
\begin{figure}
  \centering
  \input{dgmZZ.tex}
  \caption{Diagrams that contribute in the MSSM to the decay $A\to ZZ$. In the
    case of charged particles in the loop the diagrams with reversed charge flow
    have been omitted.  In the 2HDM the 1-loop decay amplitude is due to the
    diagrams (a--c) only.}
  \label{fig:dgmZZ}
\end{figure}

\begin{figure}
  \centering
  \includegraphics{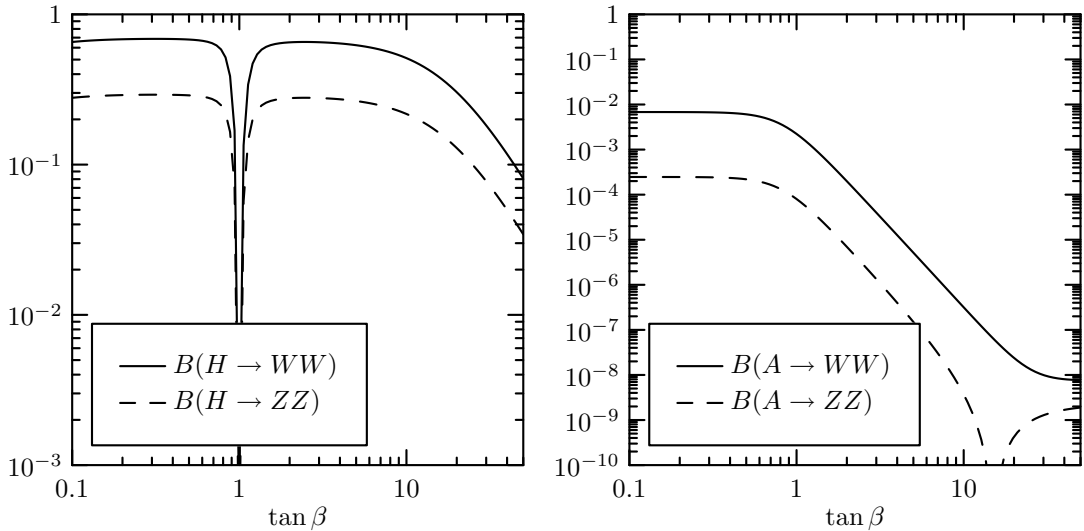}
  \caption{The branching ratios for $H\to WW,ZZ$ (left) and $A\to WW,ZZ$ (right)
    in the 2HDM as functions of $\tan\beta$ for $m_H=m_A=\unit{250}{GeV}$,
    $m_h=\unit{160}{GeV}$, and $\alpha=-\pi/4$.}
  \label{fig:tb-BHH.thdm}
\end{figure}

We consider here a type-II two-Higgs doublet extension (2HDM)
\cite{Gunion:1989we} of the SM (with CP-invariant tree-level Higgs potential),
where by construction flavour-changing neutral current interactions are absent
at tree level. The Higgs sector of this model involves, besides the masses of
the charged and the three neutral Higgs bosons, three more free parameters,
which are commonly denoted as $\tan\beta$, $\alpha$ and $\lambda_5$.  The
parameter $\tan\beta$ is the ratio ${v}_u/{v}_d$, where ${v}_{u,d}$ are the
vacuum expectation values of the neutral components of the Higgs doublet fields
$\Phi_{u,d}$. The angle $\alpha$ describes the mixing of the two $CP$-even
neutral Higgs states which leads to the mass eigenstates $h$ and $H$. The
parameter $\lambda_5$ is a dimensionless coupling from the term
\begin{equation}
  \lambda_5 (\tfrac12(\Phi_d^\dagger \Phi_u + \Phi_u^\dagger\Phi_d) - v_u v_d)^2
\end{equation}
in the Higgs potential.\footnote{See \cite{Gunion:1989we} for details.}
 
The decays $A\to WW$ and $A\to ZZ$ are mediated by the 1-loop diagrams (a-c) of
Fig.~\ref{fig:dgmWW} and Fig.~\ref{fig:dgmZZ}, respectively.  For computing the
branching ratios $B(A, H \to WW/ZZ)$, we need to determine the total widths of
$A$ and $H$. For this purpose the following decays should in addition be taken
into account, if kinematically possible:
\begin{align}\label{eq:res:thdm-decays}
  H& \to  W^\pm H^\mp, \ hh,\ b\bar b,\ \tau \bar\tau, \ t\bar t, \ gg,
  \nonumber\\
  A& \to  W^\pm H^\mp, \ Z h, \ b\bar b,\ \tau\bar\tau,\ t\bar t, \ gg
  \eqpunct.
\end{align} 
Apart from the $gg$ mode, these decays are induced already at tree level. The
$H,A \to gg$ decay amplitudes are given by diagrams analogous to
Fig.~\ref{fig:dgmZZ} (b,c) and are dominated by the top-quark loop for small and
moderate values of $\tan\beta$. For definiteness we put the mass of the light
scalar Higgs boson 
\begin{equation}
 m_h= 160 \, {\rm GeV} \, .
\end{equation}

As our goal is to investigate how large the branching ratios $B(A\to WW/ZZ)$ and
the ratios $R_W, R_Z$ defined in (\ref{rvratio}) can become, we choose the mass
of $H^\pm$ such that the decays $A\to W^\pm H^\mp$ are kinematically
forbidden. (In fact, a charged Higgs boson with a mass below 315 GeV is excluded
in the 2HDM on empirical grounds \cite{Gambino:2001ew}.)

First we consider a relatively light pseudoscalar $A$ and, for comparison, a
scalar $H$ of the same mass: $m_A=m_H=250$ GeV. Then, apart from $A,H\to W^\pm
H^\mp$, also the channels $A\to Zh, t{\bar t}$ and $H\to hh, t{\bar t}$ are
closed in view of the chosen Higgs-boson masses.  The resulting branching ratios
for $H,A\to WW/ZZ$ are shown as functions of $\tan\beta$ in
Fig.~\ref{fig:tb-BHH.thdm}, for a fixed mixing angle $\alpha=-\pi/4$.
 For small $\tan\beta$ the total $H$ decay width is
dominated by the decays $H\to WW, ZZ$. (The $gg$ mode contributes about 6$\%$
for $\tan\beta=0.1$.)  As the Born amplitudes of both modes are proportional to
$\cos(\beta-\alpha)$ the branching ratios $B(H\to VV)$ are almost constant in
that region.  For $|\beta-\alpha|=\pi/2$, i.e., $\tan\beta=1$ for our choice of
$\alpha$, the Born amplitudes vanish, which causes the sharp dip in
Fig.~\ref{fig:tb-BHH.thdm} (left panel). The one-loop corrections render the
amplitude non-zero, and we expect $B(H\to VV) \approx B(A\to VV)$ in that region
of parameter space.  For large $\tan\beta$ the width of $H\to b\bar b$ gets
bigger and eventually exceeds those of $H\to VV$. As a result the $H\to WW,ZZ$
branching ratios decrease for large $\tan\beta$.

The total $A$ decay width is dominated, for small $\tan\beta$, by the
contribution from $A\to gg$. The $A\to WW,ZZ,gg$ decay amplitudes are
essentially determined by the diagrams where $A$ couples to the top quark, i.e.,
they are are all proportional to $\cot\beta$.  Thus the branching ratios of
$A\to WW,ZZ$ are approximately constant for $\tan\beta\lesssim 1$. For
$\tan\beta>1$ the $A\to WW,ZZ,gg$ partial widths get smaller with increasing
$\tan\beta$ and the (tree-level) $A\to b\bar b$ partial width, which is
proportional to $\tan^2\beta$, begins to dominate. As a result the branching
ratios $B(A\to WW,ZZ)$ decrease for $\tan\beta>1$ in the range shown in
Fig.~\ref{fig:tb-BHH.thdm}.

\begin{figure}
  \centering
  \includegraphics{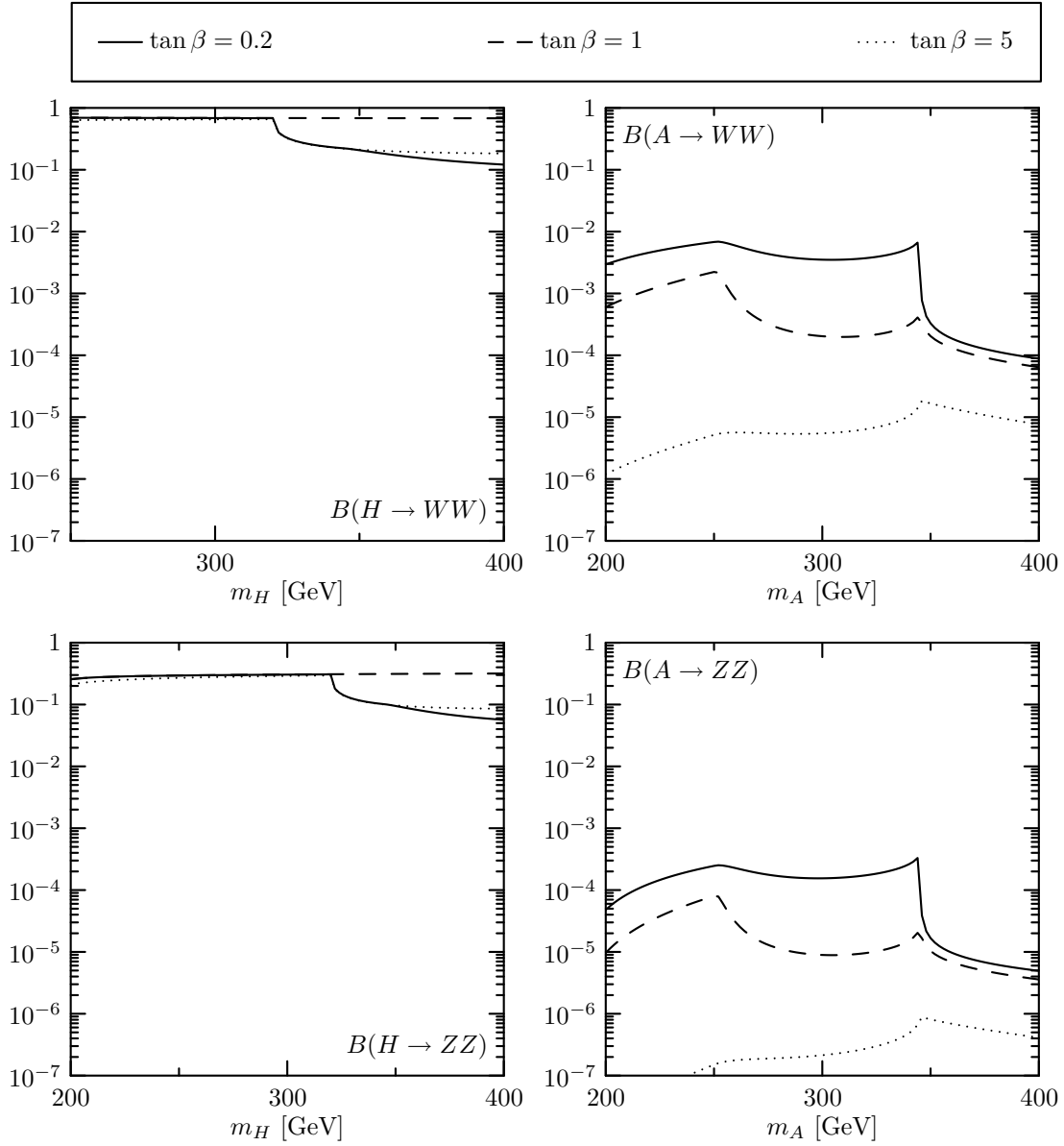}
  \caption{The branching ratios for $H\to WW,ZZ$ (left) and $A\to WW,ZZ$ (right)
    in the 2HDM as functions of $m_H$ and $m_A$, respectively, for three
    different values of $\tan\beta$.  The mixing angle $\alpha$ is chosen such
    that $\alpha=\beta -\pi/4$, and $m_h=\unit{160}{GeV}$.}
  \label{fig:mA0.thdm}
\end{figure}

In Fig.~\ref{fig:mA0.thdm} the branching ratios of $H,A\to WW/ZZ$ are shown as
functions of $m_{H,A}$ for three different values of $\tan\beta$. In order to
exclude the region around $|\beta-\alpha|=\pi/2$ we put $\alpha=\beta
-\pi/4$. The mass of $h$ is again chosen to be $m_h= 160$ GeV.
Fig.~\ref{fig:mA0.thdm} shows that, once the decay channel $H\to hh$ is open, the
branching ratios $B(H\to VV)$ decrease significantly if $\tan\beta$ is markedly
smaller or larger than one, because in a generic 2HDM the $H\to hh$ amplitude is
proportional to $1/\sin2\beta$, so that $H\to hh$ dominates the total width
$\Gamma_H$, if $\tan\beta$ is small or large. Note, however, that the $Hhh$
vertex also contains a term proportional to $\lambda_5$. Here we set
\begin{equation}
  \lambda_5=0
  \eqpunct,
\end{equation}
but it is also possible to tune $\lambda_5$ to cancel the enhancement factor
$1/\sin2\beta$. In fact, this is precisely what happens in the MSSM, where
$\lambda_5$ is no longer an independent parameter.

As far as $A$ is concerned, the opening of $A\to Zh$ (the amplitude of which is
proportional to $\cos(\beta-\alpha)$) has a less dramatic effect on the $A\to
VV$ branching ratios. Once $A\to t\bar t$ is kinematically possible, they drop,
however, significantly if $\tan\beta$ is not very large.

\begin{figure}
  \centering
  \includegraphics{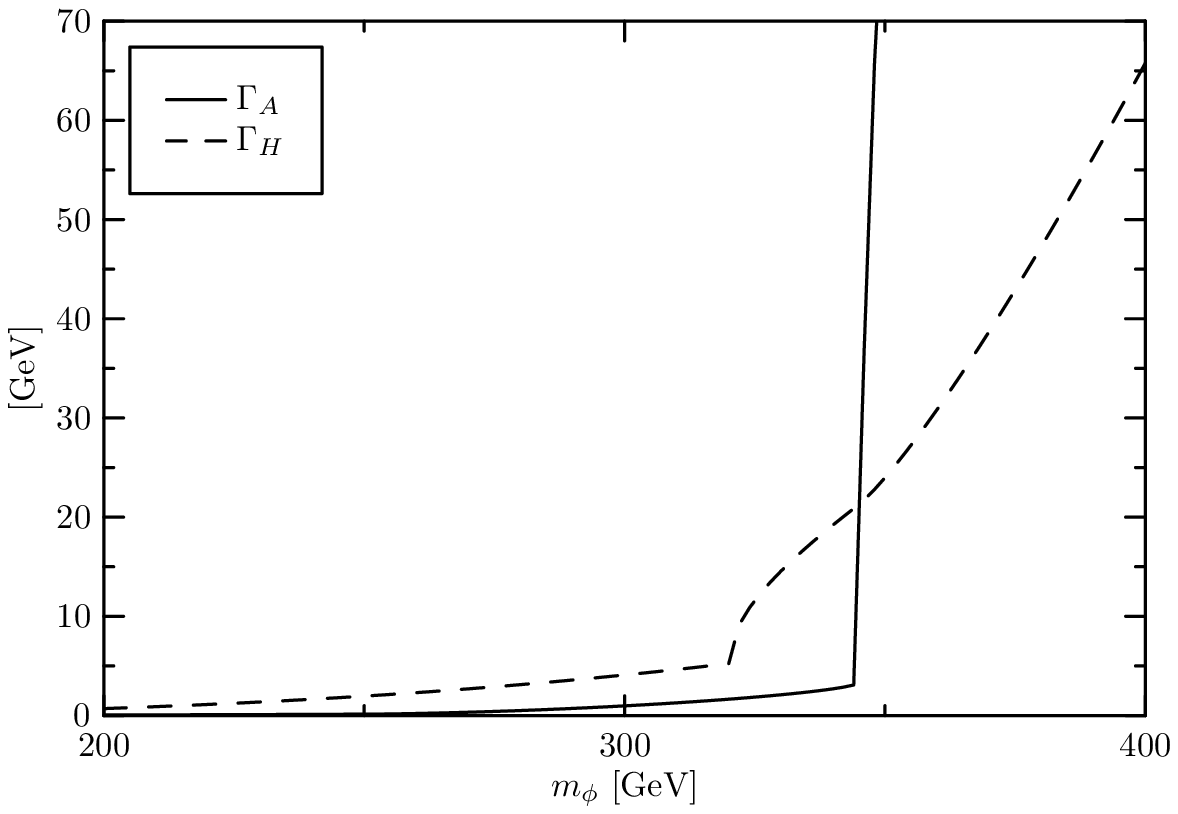}
  \caption{The  total widths of $H$ and $A$ in the 2HDM as functions
    of $m_\phi$ ($\phi=H,A$) for
    $\tan\beta=0.2$, $\alpha=\beta -\pi/4$ and $m_h=\unit{160}{GeV}$.}
  \label{fig:mA0-Gam.thdm}
\end{figure}
For completeness we show in Fig.~\ref{fig:mA0-Gam.thdm} the total widths of
$A$ and $H$ as functions of $m_\phi$ ($\phi=H,A$) for $\tan\beta=0.2$,
$\alpha=\beta -\pi/4$ and $m_h=\unit{160}{GeV}$. We see that the $H$ total width
increases rapidly when the $H\to hh$ channel opens. As pointed out
above, this is due to the
$1/\sin2\beta$ enhancement of the $Hhh$ vertex.
   The total width of  $A$ increases even more
dramatically when the $A\to t\bar t$ channel opens. This only happens for small
values of $\tan\beta$, where the $At\bar t$ coupling is enhanced. In fact, for
large $\tan\beta$ and $m_A>2m_t$ the width $\Gamma_A$  very quickly gets as
large as $m_A$ so that the pseudoscalar Higgs would no longer be visible as a
resonance.

We conclude that in the 2HDM the branching ratios $B(A\to WW)\lesssim 10^{-2}$ and
$B(A\to ZZ)\lesssim 10^{-3}$. In the case of the $WW$ final state the percent level can
be reached if $m_A \leq 2m_t$.  The corresponding branching ratios of the heavy
scalar Higgs boson $H$ with a mass $2 m_V < m_H < 2m_t$ can be quite small,
$B(H\to WW)\gtrsim 6\times 10^{-2}$ and $B(H\to ZZ)\gtrsim 3\times 10^{-2}$, if
$\tan\beta$ is significantly smaller or larger than one and $H\to hh$ is
kinematically possible and enhanced. Then this mode dominates the total width $\Gamma_H$,
which gets rather large.  In this case and for parameters as in
Fig.~\ref{fig:mA0.thdm} the ratios (\ref{rvratio}) are $R_W \lesssim 0.17$ and
$R_Z < 0.03$. However, in a small region of parameter space, in the close
vicinity of $|\beta-\alpha|=\pi/2$, these ratios can be even of order one.
% 
%
%%%%%%%%%%%%%%%%%%%%%%%%%%%%%%%%%%%%%%%%%%%%%%%%%%%%%%%%%%%%%%%%%%%%%%%%%%%%%%%%
\subsection{The MSSM} \label{subsec:MSSM}
%%%%%%%%%%%%%%%%%%%%%%%%%%%%%%%%%%%%%%%%%%%%%%%%%%%%%%%%%%%%%%%%%%%%%%%%%%%%%%%%
%
In the MSSM the masses of the neutral Higgs bosons and the mixing angle $\alpha$
are no longer independent parameters. At tree level they can be expressed in
terms of the charged Higgs-boson mass $m_{H^\pm}$ and $\tan\beta$ as follows:
\begin{subequations}
\label{eq:tree_level_relations}
\begin{gather}
  m_A^2 = m_{H^\pm}^2 + m_W^2
  \eqpunct,\label{eq:m_A}\\
  m_{H,h}^2 = \frac12\left[m_A^2+m_Z^2\pm
  \sqrt{(m_A^2+m_Z^2)^2 - 4m_Z^2m_A^2\cos^22\beta}\right]
  \eqpunct,\\
  \cos(2\alpha) = -\cos(2\beta)\frac{m_A^2-m_Z^2}{m_H^2-m_h^2}
  \quad,\quad
  \sin(2\alpha) = -\sin(2\beta)\frac{m_H^2+m_h^2}{m_H^2-m_h^2}
  \eqpunct.
\end{gather}
\end{subequations}
For MSSM scenarios with $m_{H^\pm}\gg m_Z$ these equations yield
\begin{equation}\label{eq:approximations}
  m_A\approx m_H \quad,\quad \beta-\alpha\approx\frac\pi2
  \eqpunct.
\end{equation}
The $HWW$ and $HZZ$ Born vertices are proportional to $\cos(\beta-\alpha)$, as
in the 2HDM.  Consequently the $H\to WW$ and $H\to ZZ$ branching ratios are
naturally suppressed in MSSM scenarios with heavy Higgs particles.

However, it is well known that the tree-level relations
\eqref{eq:tree_level_relations} are substantially modified by loop corrections
to the MSSM Higgs potential. These corrections are responsible for pushing the
the mass of the light Higgs boson above the current lower limit of
$\approx\unit{115}{GeV}$ and have to be taken into account to obtain reliable
results. The lower limit on $m_h$ implies that $\tan\beta\gtrsim 3$ in the MSSM.

Here we  used \texttt{FeynHiggs} version 2.6.5 
\cite{Frank:2006yh,Degrassi:2002fi,Heinemeyer:1998np,Heinemeyer:1998yj}
 to calculate all one-loop and leading two-loop corrections to the neutral
Higgs-boson self-energies in the MSSM and extract from them the physical neutral
Higgs masses, LSZ residues, and the resulting effective mixing angle
$\alpha$. Our results for the partial widths of $H$ and $A$ were then calculated
with these effective parameters. It turns out that the approximations
\eqref{eq:approximations} for $m_{H^+}$ being large are still valid if loop
corrections to the Higgs potential are taken into account. Consequently the
factor $\cos(\beta-\alpha)$ in the $HVV$ vertices is still small. Our SUSY
parameters are chosen in such a way that the following mass bounds are
satisfied:
\begin{gather}
  m_h\gtrsim\unit{115}{GeV}
  \quad,\quad
  m_{\chi^+_1}\gtrsim\unit{100}{GeV}
  \quad,\quad
  m_{\chi^0_1}\gtrsim\unit{50}{GeV}
  \quad,\quad
  m_{\tilde t_1}\gtrsim\unit{100}{GeV}
  \eqpunct,\nonumber\\
  m_{\tilde t_2},m_{\tilde q},m_{\tilde l},m_{\tilde\nu}\gtrsim\unit{500}{GeV}
  \eqpunct.
\end{gather}
For this purpose, we fix the following soft masses and trilinear couplings:
\begin{gather}
  \mu = 200\,{\rm GeV}
  \quad,\quad
  M_1 = \frac53\tan^2\theta_W M_2
  \quad,\quad
  m_{\tilde g}=\unit{500}{GeV}
  \eqpunct,\nonumber\\
  M_{L_i}=M_{E_i}=M_{Q_i}=M_{D_i}=M_{U_{1,2}}=A_{L_i}=A_{U_{1,2}}=A_{D_i}=
    \unit{500}{GeV}
  \eqpunct.\label{eq:susy_parameters}
\end{gather}
Here $\mu$, $M_1$, and $M_2$ are the soft SUSY breaking masses of the Higgs
potential, $M_L$ and $M_E$ are the soft masses of the left- and right-handed
sleptons, $M_Q$ the soft masses of the left-handed squarks, and $M_U$ and $M_D$
the soft masses of the right-handed up- and down-type squarks. Likewise, $A_L$,
$A_U$ and $A_D$ are the trilinear couplings of the sleptons, up-type squarks and
down-type squarks, respectively. Furthermore $m_{\tilde g}$ is the gluino mass
and $i=1,2,3$ is a generation index. The relation between $M_1$ and $M_2$ is the
usual GUT relation. Note that the $A\to VV$ partial widths are
affected directly only by  the parameters
$\tan\beta$, $\mu$, $M_1$, and $M_2$, because
 the only non-SM particles appearing in the loop diagrams in
Fig.~\ref{fig:dgmWW} and~\ref{fig:dgmZZ} are the charginos and neutralinos.

We left $m_{H^\pm}$, $M_2$, $M_{U_3}$ and $A_{U_3}$ unspecified in
\eqref{eq:susy_parameters}. Rather than choosing values for these parameters
directly, we will adjust them numerically to obtain specific values for the
pseudoscalar mass $m_A$, the lightest Higgs-boson mass $m_h$, the lightest stop
mass $m_{\tilde t_1}$, and the lightest neutralino mass
$m_{\tilde\chi^0_1}$. Throughout this section we use
\begin{equation}\label{eq:m_h}
  m_h = \unit{115}{GeV}
  \quad,\quad
  m_{\tilde t_1} = \unit{400}{GeV}
  \eqpunct.
\end{equation}

The MSSM diagrams that contribute to $A\to WW$ and $A\to ZZ$ decays are shown in
figures \ref{fig:dgmWW} and \ref{fig:dgmZZ}, respectively. As we discussed in
the 2HDM, the contributions from loops involving $t$ or $b$ quarks are rather
small for $\tan\beta>1$. As a result the dominant contributions to the $A\to
WW,ZZ$ partial widths in the MSSM come from diagrams with charginos or
neutralinos in the loop, as long as their masses are not substantially larger
than $m_A$.

\begin{figure}
  \centering
  \includegraphics{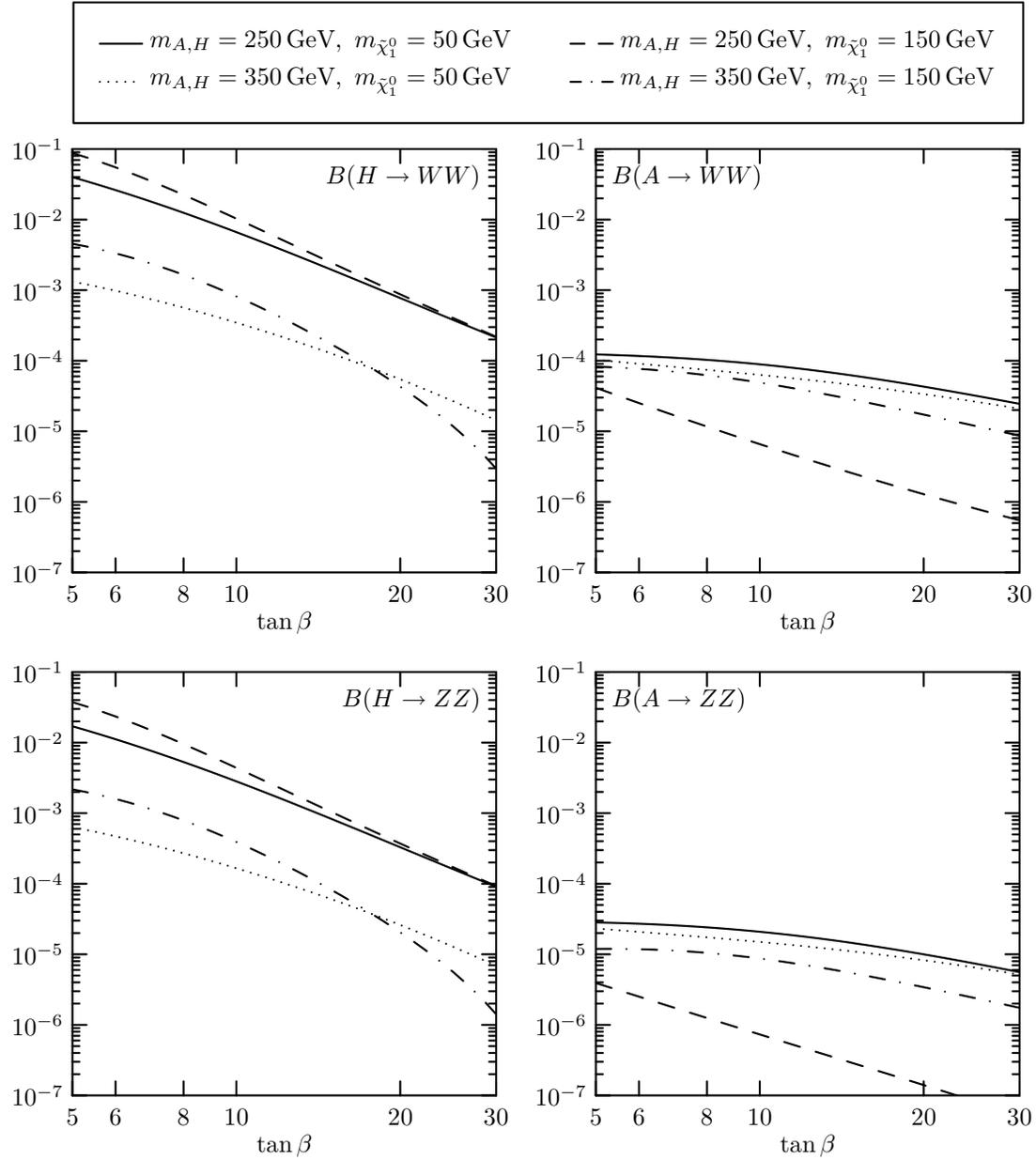}
  \caption{The MSSM branching ratios of the decays of
    $H$ (left column) and $A$ (right column) into 
     $WW$ (first row) and $ZZ$ (second row) final states as functions of
    $\tan\beta$ for the parameters \eqref{eq:sm_parameters},
    \eqref{eq:susy_parameters}, \eqref{eq:m_h}, 
   and different combinations of $m_A$ and
    $m_{\tilde\chi^0_1}$.}
  \label{fig:tb.mssm}
\end{figure}
Fig.~\ref{fig:tb.mssm} shows the branching ratios of $H,A \to WW/ZZ$ as
functions of $\tan\beta$ for the parameters \eqref{eq:sm_parameters},
\eqref{eq:susy_parameters}, \eqref{eq:m_h} and for different combinations of
$m_A$ and $m_{\tilde\chi^0_1}$.  The total widths of $A$ and $H$ were calculated
with \texttt{FeynHiggs}
\cite{Frank:2006yh,Degrassi:2002fi,Heinemeyer:1998np,Heinemeyer:1998yj}.
 In our scenario the contributing decay channels
are
\begin{align}
  H&\to\ b\bar b,\  \tau\bar\tau, \ t\bar t, \ gg,\ \chi^+_i\chi^-_j,\
        \chi^0_i\chi^0_j,\ hh
  \eqpunct,\nonumber\\
  A&\to \ b\bar b,\ \tau\bar\tau, \ t\bar t, \ gg,\ \chi^+_i\chi^-_j,\
        \chi^0_i\chi^0_j,\ Zh
  \eqpunct.
\end{align}
For light  $\tilde\chi^0_1$, $\tan\beta \gtrsim 5$, and $m_{H,A} <2 m_t$
the dominant contributions to the $A$ and $H$ total widths come from the decays
into $b\bar b$, $\tilde\chi^+_1\tilde\chi^-_1$ and
$\tilde\chi^0_2\tilde\chi^0_1$, provided the latter are kinematically allowed.  For
$\tan\beta\gtrsim 15$ the $A$ and $H$ total widths are dominated by the $b\bar
b$ ($\sim 90\%$) and $\tau^-\tau^+$ ($\sim 10\%$) channels, as long as the
$t\bar t$ channel is closed.  In the $\tan\beta$ range shown in
Fig.~\ref{fig:tb.mssm} the total widths of $A$ and $H$ grow by approximately one
order of magnitude. 

At the same time the $H\to WW,ZZ$ partial widths drop by approximately two
orders of magnitude because $\cos(\beta-\alpha)$ becomes
small. Nevertheless, for the parameters used in Fig.~\ref{fig:tb.mssm},
the one-loop contributions to $H\to VV$ are still smaller than the
Born terms.  For large $\tan\beta$ and large $m_{A,H}$ the $H\to WW,ZZ$ branching ratios can
actually be smaller than the corresponding $A$ branching ratios. 

The $A\to WW,ZZ$ partial widths and hence 
 the $A\to WW,ZZ$ branching ratios are less sensitive to $\tan\beta$.
However, $B(A\to WW)<10^{-4}$ and $B(A\to ZZ) <2\times 10^{-5}$ for the
parameters used in  Fig.~\ref{fig:tb.mssm}.

\begin{figure}
  \centering
  \includegraphics{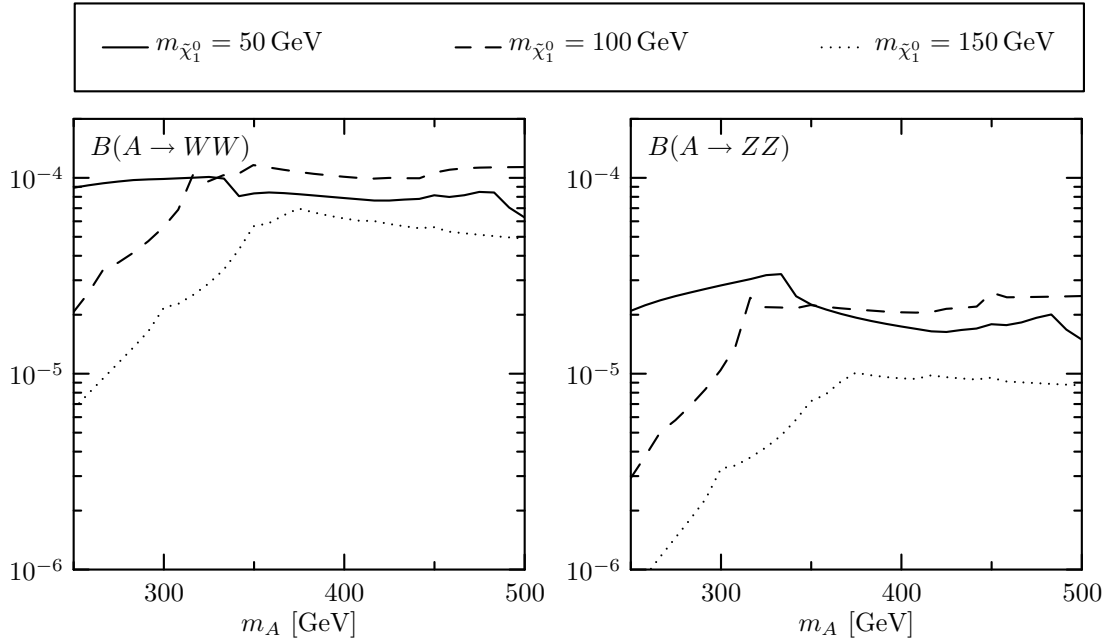}
  \caption{The branching ratios for $A\to WW$ (left) and $A\to ZZ$ (right) as
    functions of $m_A$ for the parameters \eqref{eq:sm_parameters},
    \eqref{eq:susy_parameters}, \eqref{eq:m_h}, $\tan\beta=10$, and three
    different values of $m_{\tilde\chi^0_1}$.}
  \label{fig:mA0.mssm}
\end{figure}
In Fig.~\ref{fig:mA0.mssm} the branching ratios of $A\to WW, ZZ$ and $A\to ZZ$
are shown as functions of $m_A$ for different values of
$m_{\tilde\chi^0_1}$. The kinks correspond to the opening of decay channels into
chargino or neutralino pairs. The drop in the branching ratio that is usually
associated with the opening of a new channel is less pronounced here, because
whenever the total width increases due to the opening of such a channel, a
corresponding loop diagram from Fig.~\ref{fig:dgmWW} or~\ref{fig:dgmZZ} develops
an imaginary part, which leads to an increase in the $A\to WW,ZZ$ partial
widths.

\begin{figure}
  \centering
  \includegraphics{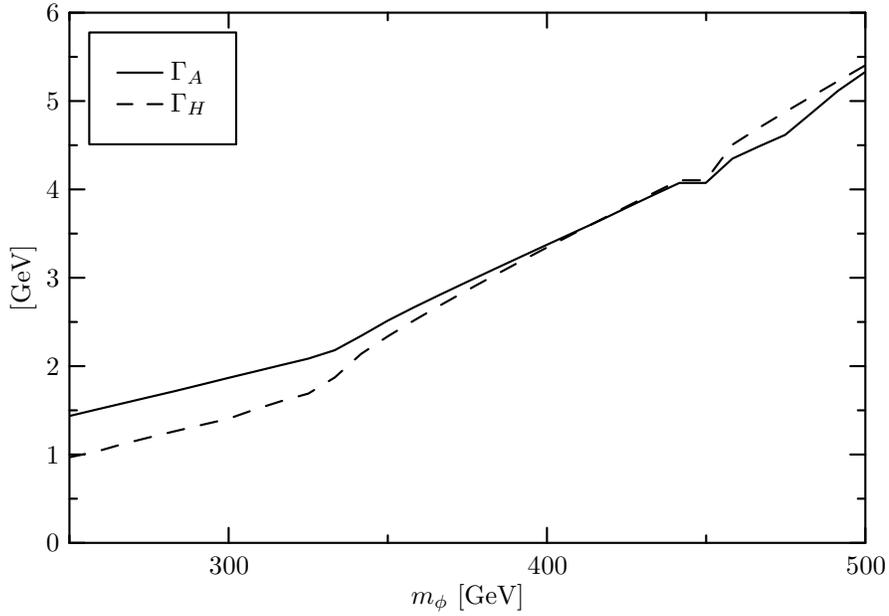}
  \caption{The $H$ and $A$ total widths in the MSSM as functions of
    $m_\phi$ ($\phi=H,A$) for
    the parameters \eqref{eq:sm_parameters}, \eqref{eq:susy_parameters},
    \eqref{eq:m_h}, $\tan\beta=10$ and $m_{\tilde\chi^0_1}=\unit{50}{GeV}$.}
  \label{fig:mA0-Gam.mssm}
\end{figure}
Fig.~\ref{fig:mA0-Gam.mssm} shows the $H$ and $A$ total widths as functions of
their mass for the parameters \eqref{eq:sm_parameters},
\eqref{eq:susy_parameters}, \eqref{eq:m_h}, $\tan\beta=10$ and
$m_{\tilde\chi^0_1}=\unit{50}{GeV}$. The rapid increase of $\Gamma_H$ at the
opening of the $H\to hh$ threshold is absent here, because the MSSM constraints
on the 2HDM Higgs potential parameters tune $\lambda_5$ in such a way that the
enhancing factor $1/\sin2\beta$ cancels in the $Hhh$ coupling. Since $\tan\beta$
is large here, the opening of the $A\to t\bar t$ threshold also has no dramatic
effect on the $A$ total width. As a result, the $A$ and $H$ total widths stay
below \unit{6}{GeV} in the whole mass range $\unit{250}{GeV}\leq
m_\phi\leq\unit{500}{GeV}$.

From Figs.~\ref{fig:tb.mssm} and~\ref{fig:mA0.mssm} we conclude that in the MSSM
the decays $A\to WW,ZZ$ are really rare, as the branching ratios $B(A\to
WW)\lesssim 10^{-4}$ and $B(A\to ZZ) \lesssim 2\times 10^{-5}$.  The branching
ratios of the corresponding decays of the heavy scalar Higgs boson $H$ depend
sensitively on the MSSM parameters, in particular on the value of
$\tan\beta$. In the scenario $m_{H^\pm}, m_H, m_A \gg m_Z$ the branching ratios
$B(H\to WW,ZZ)$ are, for $m_H=250$ GeV, about 0.09 and 0.04, respectively, for
$\tan\beta=5$, but drop below $10^{-3}$ for large $\tan\beta$.
%
%
%%%%%%%%%%%%%%%%%%%%%%%%%%%%%%%%%%%%%%%%%%%%%%%%%%%%%%%%%%%%%%%%%%%%%%%%%%%%%%%%
\subsection{A heavy fourth generation}
%%%%%%%%%%%%%%%%%%%%%%%%%%%%%%%%%%%%%%%%%%%%%%%%%%%%%%%%%%%%%%%%%%%%%%%%%%%%%%%%
%
Recently, there has been renewed interest in the question of whether a
sequential fourth generation of chiral quarks and leptons (left-chiral doublets
and right-chiral singlets) with masses in the (few) hundred GeV range can exist,
in spite of strong experimental constraints (see, e.g.,
\cite{Kribs:2007nz,Holdom:2009rf}). In the following we shall denote the
corresponding fourth generation quark and lepton mass eigenstates by $u_4, d_4,
\ell_4, \nu_4$.  The strongest constraints on the masses of these states come
from direct searches at LEP2 and at the Tevatron, and from electroweak precision
observables. Non-observation at LEP2 implies the lower bounds $m_{\ell_4} >100$
GeV, $m_{\nu_4} > 90$ GeV (for definiteness, we assume $\nu_4$ to be a Dirac
particle) \cite{Amsler:2008zzb}, while searches for heavy quarks at the Tevatron
yield the mass limits $m_{d_4} > 190$ GeV \cite{Amsler:2008zzb} and $m_{u_4} >
311$ GeV \cite{Eusebi:2009sh}. Strong constraints on a fourth generation are
also implied by the oblique electroweak corrections, i.e., by the experimentally
allowed values of the $S,T$ and $U$ parameters. While the mass splitting of the
$I_W=\pm 1/2$ partners of a doublet need not be too large in order not to
violate the bound on the $\rho$ parameter respectively on $T$, degenerate
doublets are not acceptable, too, because this would make the contribution
$\delta S$ to the $S$ parameter too big. However, if $m_{u_4}> m_{d_4}$ and
$m_{\ell_4} > m_{\nu_4}$ then $\delta S$, $\delta T$ can be brought in accord
with the experimental constraints.
 
In the following we consider a type-II 2HDM as in Sect.~\ref{sec:res:THDM} with
four sequential quark and lepton generations.  For our numerical analysis we use
the following masses for the fourth generation fermions:
\begin{equation}
 m_{d_4} =
\unit{200}{GeV}, \, m_{u_4} = \unit{320}{GeV}, \, m_{\ell_4} = \unit{220}{GeV},
\, m_{\nu_4} = \unit{180}{GeV}.
\end{equation}
 As in Sect.~\ref{sec:res:THDM} we put
$m_h=\unit{160}{GeV}$ and $\beta-\alpha=\pi/4$.

\begin{figure}
  \centering
  \includegraphics{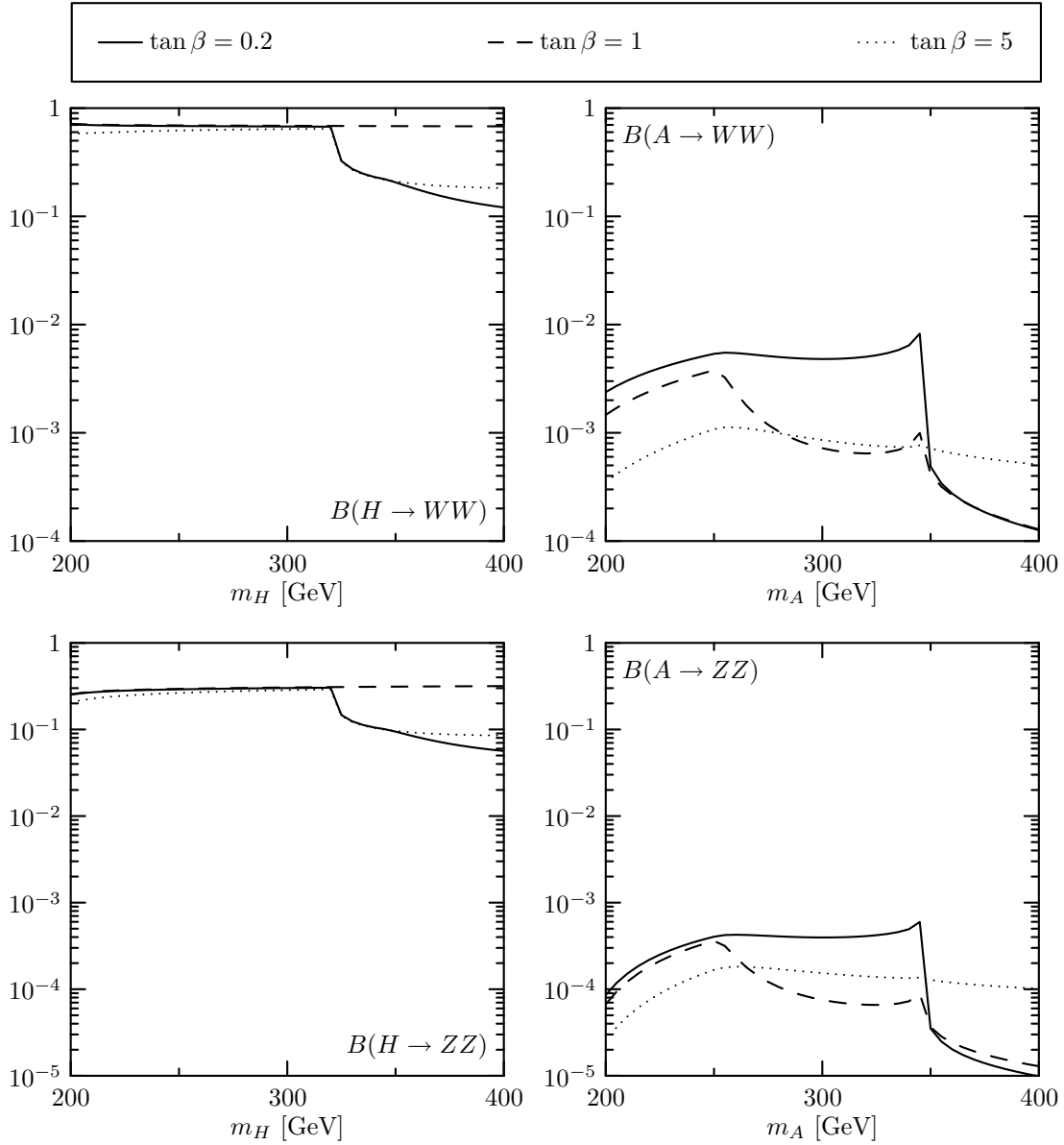}
  \caption{\label{fig:thdm4} Heavy Higgs-boson branching ratios in the 2HDM with a
    fourth generation of chiral fermions. Shown are the branching ratios of $H$
    (left column) and $A$ (right column) into $WW$ (first row) and $ZZ$ (second
    row) as functions of the decaying particle's mass for different values of
    $\tan\beta$. The remaining parameters are as given in the text. }
\end{figure}%

Fig.~\ref{fig:thdm4} shows the $H$ and $A$ branching ratios into $WW$ and $ZZ$
as functions of the decaying particle's mass for different values of
$\tan\beta$. The decays that contribute to the total widths are the same
as in \eqref{eq:res:thdm-decays}, because we keep $m_{H,A}$
below the production thresholds for the 4th generation fermions. 

In order to discuss the contribution $\Delta C_{VV}$ of the fourth generation
fermions to the amplitude $C_{VV}$ of the decay $A\to VV$ (see
\eqref{def:cvv}), we may evaluate $\Delta C_{VV}$ in the limit where the masses
of these fermions are much larger than the masses of the external particles.
Adding up the contributions of $u_4, d_4, \ell_4,$ and $\nu_4$ we obtain:
\begin{align}
  \Delta C_{WW} &= \frac{i\alpha}{6\pi s_W^2}(\tan\beta+\cot\beta)
  \eqpunct,\label{eq:res:CWW}\\
  \Delta C_{ZZ} &= \frac{i\alpha}{6\pi s_W^2}
           (1-3s_W^2+4s_W^4)(\tan\beta+\cot\beta)
  \eqpunct.
  \label{eq:res:CZZ}
\end{align}
For the masses chosen above and in Fig.~\ref{fig:thdm4} , these formulae are in
fact reasonable approximations to the exact one-loop amplitudes. That is, the values
for $B(A,H \to VV)$ shown in Fig.~\ref{fig:thdm4} apply also to much heavier
fourth-generation fermions.

Eq.~\ref{eq:res:CWW} and~\ref{eq:res:CZZ} imply that $u_4, d_4, \ell_4, \nu_4$
make a significant contribution to the partial widths $\Gamma(A\to VV)$,
especially for small or large $\tan\beta$. These widths are larger than those in
the three-generation 2HDM of Sect.~\ref{sec:res:THDM}.  On the other hand, $u_4$ and
$d_4$ increase also the partial width $\Gamma(A\to gg)$ as compared to the
three-generation 2HDM, which dominates the total width $\Gamma_A$ for small
$\tan\beta$. For large $\tan\beta$ and the above choice of masses, $A\to b{\bar
  b}$ makes the most significant contribution to $\Gamma_A$, as is the case in
the three-generation 2HDM. Thus the branching ratios $B(A\to VV)$ shown in
Fig.~\ref{fig:thdm4} are, for small $\tan\beta$, of similar size as those in the
three-generation 2HDM shown in Fig.~\ref{fig:mA0.thdm}, but are much larger than
the corresponding ones in Fig.~\ref{fig:mA0.thdm} for large $\tan\beta$.

The branching ratios of $H\to VV$ lie between $20\%$ and $70\%$ for $m_H \leq 2
m_h$.  For extreme values of $\tan\beta$ they decrease rapidly when the $H\to
hh$ decay channel opens. As in the three-generation 2HDM this feature relies on
the fact that the $Hhh$ vertex is enhanced by $1/\sin2\beta$. For large
$\tan\beta$ the $H\to VV$ branching ratios drop again by a factor of 2 when the
$H\to t\bar t$ channel opens, since this channel is enhanced for large
$\tan\beta$. The behaviour of the total widths at these thresholds is
essentially the same as in the 3-generation 2HDM shown in  Fig.~\ref{fig:mA0-Gam.thdm}.

In summary, the $A$ branching ratios do not exceed 
 $10^{-2}$ for the $WW$ and 
$10^{-3}$ for the $ZZ$ final state. The ratios $R_W$ and $R_Z$ can reach $10\%$
 and $1\%$, respectively, if the $H\to hh$ channel is open and 
  enhanced but the decay $A\to t\bar t$ is kinematically
forbidden.
%
%
%%%%%%%%%%%%%%%%%%%%%%%%%%%%%%%%%%%%%%%%%%%%%%%%%%%%%%%%%%%%%%%%%%%%%%%%%%%%%%%%
\subsection{Vector-like Quarks}
%%%%%%%%%%%%%%%%%%%%%%%%%%%%%%%%%%%%%%%%%%%%%%%%%%%%%%%%%%%%%%%%%%%%%%%%%%%%%%%%
%
\label{sec:res:thdmvq}
A more exotic possibility for new heavy fermions are vector-like quarks, i.e.,
quarks whose left- and right-chiral components have equal gauge charges (see,
e.g. \cite{Frampton:1999xi,Aguila00,delAguila:2000aa,AguilarSaavedra:2009es}).
Such states are present in a number of SM extensions, including extra
dimensional models with bulk fermions \cite{Appelquist:2000nn} and Little Higgs
models \cite{ArkaniHamed:2002qy}. Quarks of this type may exist with masses as
low as a few hundred GeV; electroweak precision observables imply only very mild
constraints \cite{Cynolter:2008ea}.

Here we extend the three-generation 2HDM by multiplets of vector-like quarks
along the lines of \cite{Aguila00}. We add a $SU(2)_L$ doublet $Q=(U,D)$ with
hypercharge $1/6$ and two singlets $D'$ and $U'$ with hypercharges $2/3$ and
$-1/3$, respectively. The couplings of the vector-quarks to the electroweak
gauge bosons are given by the $SU(2)_L\times U(1)_Y$-invariant Lagrangian
\begin{equation}
  \Lcal_{\text{VQ,gauge}} = 
  \bar Qi\Dslash Q + \bar U'i\Dslash U' + \bar D'i\Dslash D'
  - M_Q\bar QQ - M_{U'}\bar U'U' - M_{D'}\bar D'D'
  \eqpunct,
\end{equation}
where $\Dslash$ denotes the covariant derivative.
  The explicit mass terms are allowed because the left- and right-handed components
of the vector-quarks transform under the same representation of the electroweak
gauge group. We construct the Yukawa interactions of the vector-like quarks
 with two Higgs
doublets $\Phi_u=(\phi^+_u,\phi^0_u)$ and $\Phi_d=(\phi^+_d,\phi^0_d)$
in analogy to the chiral
quark sector of the type-II 2HDM, i.e., 
\begin{equation}
  \Lcal_{\text{VQ,Yuk}} = 
  -y_{U'}\bar Q_L\Phi_u^\cc U'_R
  -y_{D'}\bar Q_L\Phi_d D'_R
  -\tilde y_{U'}\bar Q_R\Phi_u^\cc U'_L
  -\tilde y_{D'}\bar Q_R\Phi_d D'_L
  +\text{h.c.}
  \eqpunct,
\end{equation}
with $\Phi_{u,d}^\cc=i\sigma_2\Phi_{u,d}^*$.  The Yukawa couplings $y_{U'}$,
$y_{D'}$, $\tilde y_{U'}$ and $y_{D'}$ can, in general, be
complex\footnote{Three of the complex phases can be absorbed by redefining the
  vector-quark fields $Q$, $U'$ and $D'$. The remaining phase may be absorbed
  into one of the Higgs doublets, but then it introduces, in general, CP
  violation in the Higgs sector.}. As we are not interested here in CP-violating
effects, we take all Yukawa couplings to be real.  Then observables are
invariant under the following parameter transformations:
\begin{align}\label{eq:res:y-transformations}
  (y_{U'}, \tilde y_{U'}, y_{D'}, \tilde y_{D'})
  &\to (-y_{U'}, -\tilde y_{U'}, y_{D'}, \tilde y_{D'})
  \eqpunct,\nonumber\\
  (y_{U'}, \tilde y_{U'}, y_{D'}, \tilde y_{D'})
  &\to (y_{U'}, \tilde y_{U'}, -y_{D'}, -\tilde y_{D'})
  \eqpunct,\nonumber\\
  (y_{U'}, \tilde y_{U'}, y_{D'}, \tilde y_{D'})
  &\to (\tilde y_{U'}, y_{U'}, \tilde y_{D'}, y_{D'})
  \eqpunct.
\end{align}
Invariance of our model under the first two transformations can be shown by
absorbing the signs into the definitions of the singlet fields $U'$ and $D'$,
respectively, while invariance under the last transformation follows from
redefining the vector-quark fields by their parity-transformed counterparts
($U\to\gamma^0U$, etc.).

When the Higgs doublets acquire vacuum expectation values
$\langle\Phi_u\rangle=(0,v_u)$ and $\langle\Phi_d\rangle=(0,v_d)$ we obtain the
following mass terms for the vector-quarks:
\begin{align}
  \Lcal_{\text{VQ,mass}} &=
  {}-y_{U'}v_u\bar U_L U'_R
  -y_{D'}v_d\bar D_L D'_R
  -\tilde y_{U'}v_u\bar U'_LU_R
  -\tilde y_{D'}v_d\bar D'_LD_R
  \nonumber\\
  &\phantom{{}={}}
  -M_Q\bar U_LU_R
  -M_Q\bar D_LD_R
  -M_{U'}\bar U'_LU'_R
  -M_{D'}\bar D'_LD'_R
  +\text{h.c.}
  \nonumber\\
  &={}-\begin{pmatrix}\bar U_L\\ \bar U'_L\end{pmatrix}^\trans
     \begin{pmatrix}M_Q & y_{U'}v_u\\ \tilde y_{U'}v_u & M_{U'} \end{pmatrix}
     \begin{pmatrix}U_R\\ U'_R\end{pmatrix}
  \nonumber\\
  &\phantom{{}={}}  
    -\begin{pmatrix}\bar D_L\\ \bar D'_L\end{pmatrix}^\trans
     \begin{pmatrix}M_Q & y_{D'}v_d\\ \tilde y_{D'}v_d & M_{D'} \end{pmatrix}
     \begin{pmatrix}D_R\\ D'_R\end{pmatrix}
    +\text{h.c.}
  \eqpunct.
\end{align}
The mass matrices can be diagonalised by independent rotations of the left- and
right-handed components of the vector-quarks. Substituting
\begin{align}
  \begin{pmatrix}D\\ D'\end{pmatrix}_{L,R}
  &=\begin{pmatrix}\cos\varphi^D_{L,R} & -\sin\varphi^D_{L,R}\\
                    \sin\varphi^D_{L,R} &  \cos\varphi^D_{L,R} \end{pmatrix}
  \begin{pmatrix}D_1\\ D_2\end{pmatrix}_{L,R}
  \eqpunct,\nonumber\\
  \begin{pmatrix}U\\ U'\end{pmatrix}_{L,R}
  &=\begin{pmatrix}\cos\varphi^U_{L,R} & -\sin\varphi^U_{L,R}\\
                    \sin\varphi^U_{L,R} &  \cos\varphi^U_{L,R} \end{pmatrix}
  \begin{pmatrix}U_1\\ U_2\end{pmatrix}_{L,R}
  \eqpunct,
\end{align}
one sees that the mass matrices become diagonal if
\begin{gather}
  t^{U,D}_{L-R}\equiv\tan(\varphi^{U,D}_L-\varphi^{U,D}_R)
  =\frac{v_{u,d}(\tilde y_{U',D'}-y_{U',D'})}{M_Q+M_{U',D'}}
  \eqpunct,\nonumber\\
  t^{U,D}_{L+R}\equiv\tan(\varphi^{U,D}_L+\varphi^{U,D}_R)
  =\frac{v_{u,d}(\tilde y_{U',D'}+y_{U',D'})}{M_Q-M_{U',D'}}
  \eqpunct.\label{eq:res:vq-mixing-angles}
\end{gather}
The mass eigenvalues are
\begin{align}\label{eq:res:vq-masses}
  m_{U_{1,2}} &= \frac12\left[
     \sqrt{(M_Q+M_{U'})^2 + v_u^2(y_{U'}-\tilde y_{U'})^2}
    \pm\sqrt{(M_Q-M_{U'})^2 + v_u^2(y_{U'}+\tilde y_{U'})^2}
  \right]
  \eqpunct,\nonumber\\
  m_{D_{1,2}} &= \frac12\left[
     \sqrt{(M_Q+M_{D'})^2 + v_d^2(y_{D'}-\tilde y_{D'})^2}
    \pm\sqrt{(M_Q-M_{D'})^2 + v_d^2(y_{D'}+\tilde y_{D'})^2}
  \right]
  \eqpunct.
\end{align}
The couplings of the pseudoscalar Higgs boson $A$ and of the $W$ and $Z$ bosons
to the mass eigenstate vector-quarks $U_{1,2}$, $D_{1,2}$ are given 
 in appendix \ref{sec:thdmvq-rules}.

For the $A\to WW,ZZ$ partial widths there are several suppression mechanisms at
work here. First note that for $y_{U',D'}=\pm\tilde y_{U',D'}$ the
Yukawa interactions
of the  vector-quarks  become 
parity-conserving\footnote{If  $y_{U',D'}=+\tilde y_{U',D'}$ this is
  immediately obvious. If $y_{U',D'}=-\tilde y_{U',D'}$ the minus signs
  can be absorbed into  $U'_R$ and $D'_R$.} and, therefore, they do not
contribute to $A\to WW,ZZ$ at one-loop.
If both
$\tilde y_{U',D'}-y_{U',D'}$ and $\tilde y_{U',D'}+y_{U',D'}$ are large,
\eqref{eq:res:vq-mixing-angles} tells us that either $\varphi^{U,D}_L$ or
$\varphi^{U,D}_R$ approaches $0$ or $\pi/2$. Then many 
 (but not necessarily all)
diagrams of Fig.~\ref{fig:res:vqWW} and~\ref{fig:res:vqZZ}
     get suppressed by sine or cosine factors of the mixing angles in the
couplings.

\begin{figure}
  \centering
  \input{vqWW.tex}
  \caption{Vector-quark contributions to the $A\to WW$ decay amplitude. The
    indices $i$, $j$ and $k$ run from 1 to 2.}
  \label{fig:res:vqWW}
\end{figure}
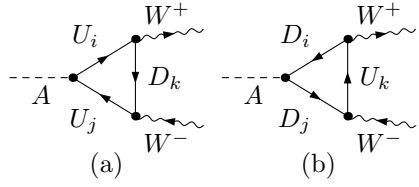
\begin{figure}
  \centering
  \input{vqZZ.tex}
  \caption{Vector-quark contributions to the $A\to ZZ$ decay amplitude. The
    indices $i$, $j$ and $k$ run from 1 to 2.}
  \label{fig:res:vqZZ}
\end{figure}
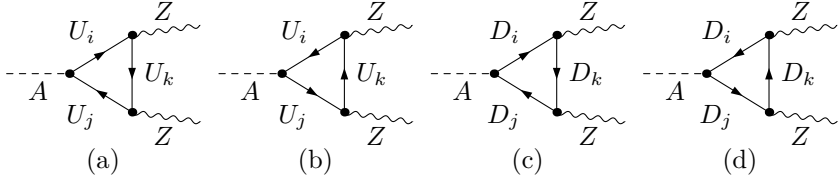

The  vector-quark contributions to
  $A\to WW$ and $A\to ZZ$ 
are shown in Fig.~\ref{fig:res:vqWW} and \ref{fig:res:vqZZ},
respectively. In
addition to the suppression mechanisms discussed above there are other
cancellations. While the
``diagonal'' couplings between A and the vector-quarks ($AU_1\bar U_1$,
$AD_1\bar D_1$ etc.)  are pure  pseudoscalar, the
``off-diagonal'' couplings ($AU_1\bar U_2$, $AD_1\bar D_2$ etc.) 
 have scalar and
pseudoscalar Lorentz structure. Yet the contributions of the
scalar terms to the form factor $f_{3,VV'}$ from \eqref{phivv}
must cancel, and this is 
 seen by observing that
 these terms change sign if the fermion flow in the loop is reversed. 
 There are further cancellations
between the pseudoscalar terms. For example, for $M_Q=M_{U'}=M_{D'}$ 
  the $A\to ZZ$ diagrams with only one type of vector-quark in the loop
cancel exactly in the limit of large $M_Q$.

To control the various suppression mechanisms, we invert
\eqref{eq:res:vq-mixing-angles} and \eqref{eq:res:vq-masses} and use the
following quantities as independent parameters:
\begin{equation}\label{eq:res:thdmvq-param}
  M_Q\ ,\ t^U_{L-R}\ ,\ t^U_{L+R}\ ,\ m_{U_2}\ ,\
          t^D_{L-R}\ ,\ t^D_{L+R}\ ,\ m_{D_2}
  \eqpunct.
\end{equation}
Using the invariance under the transformations \eqref{eq:res:y-transformations}
we may limit ourselves to the case
\begin{equation}
  t^U_{L-R},\, t^D_{L-R}, \, t^D_{L+R} > 0
  \eqpunct.
\end{equation}
Note that $t^U_{L+R}$ can still be negative.
%  By choosing the lightest physical
% vector-quark masses as input parameters we can control the suppression by heavy
% particles in the loops. Suppressions by sine or cosine factors appear when all
% $t$s get large while suppression by restored parity symmetry appears if one of
% the $t$s vanishes.

Let us now come to the numerical analysis. 
 Throughout the following discussion we
set
\begin{equation}\label{eq:res:thdmvq-alpha-MQ}
  \beta-\alpha = \frac\pi4
  \quad,\quad
  M_Q = \unit{1}{TeV}
  \eqpunct.
\end{equation}
Our choice for $\alpha$ assures that the $HVV$ Born vertices  
 are not suppressed. We will first discuss a
scenario where the pseudoscalar Higgs boson is too light to decay into
vector-quarks. In this case the leading contributions to the $H$ and $A$ total
widths are again the decays  \eqref{eq:res:thdm-decays}. As in the
case of  the 2HDM
with a 4th generation the contributions of the new heavy quarks 
to the partial widths of $H\to gg$  and $A\to gg$ 
have to be taken
into account, too. 
We further distinguish between scenarios with small
or large values of $\tan\beta$. The contributions of vector-quark loops to $A\to
WW$ and $A\to ZZ$ decays generally saturate if the magnitude of the $t$
parameters (\ref{eq:res:vq-mixing-angles}) is larger than 10. The biggest corrections were found for
$t^D_{L-R},t^D_{L+R},t^U_{L-R}\gtrsim 10$ and $t^U_{L+R}\lesssim -10$. Unless
stated otherwise, we therefore choose for our numerical analysis
\begin{gather}
  m_{U_2}=m_{D_2}=\unit{320}{GeV},
  \quad
   t^D_{L-R}=t^D_{L+R}=t^U_{L-R}=10, 
  \quad
  t^U_{L+R}=-10 \, .
  \label{eq:res:thdmvq-lhparam}
\end{gather}

\begin{figure}
  \centering
  \includegraphics{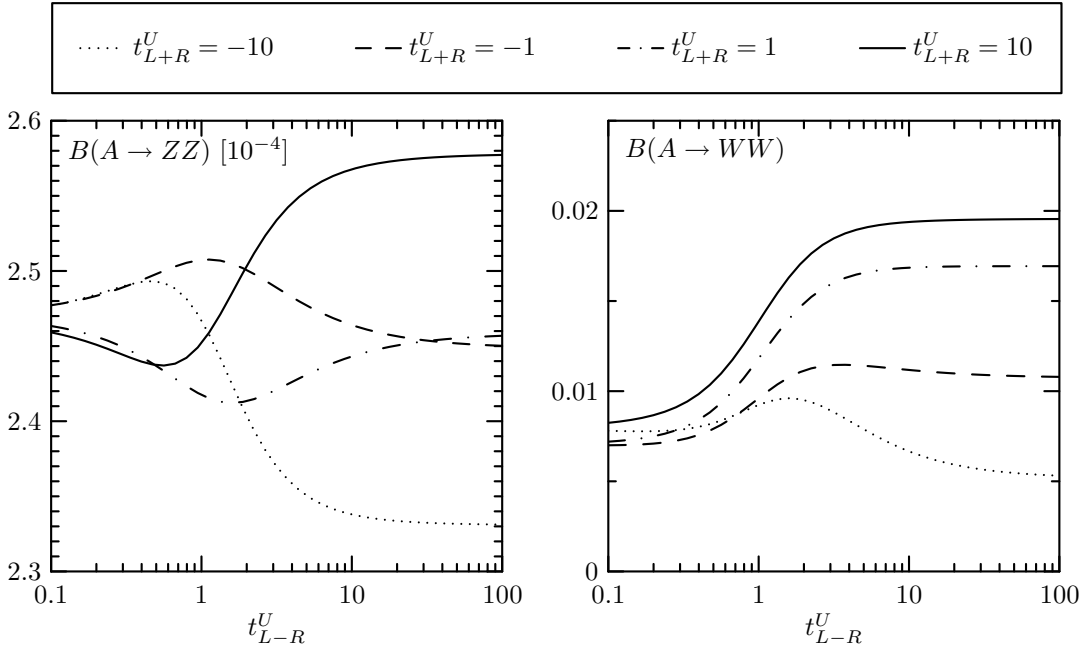}
  \caption{Results for the $A\to ZZ$ (left) and $A\to WW$ (right) branching
    ratios in the 2HDM with vector-quarks for the parameters from \eqref{eq:res:thdmvq-alpha-MQ}
    and \eqref{eq:res:thdmvq-lhparam},  $m_A=250$ GeV,
    $\tan\beta=0.2$ and different values of $t^U_{L+R}$ as functions of
    $t^U_{L-R}$.}
  \label{fig:res:tUD-thdmvq-smalltb}
\end{figure}

Fig.~\ref{fig:res:tUD-thdmvq-smalltb} shows, for $m_A=250$ GeV, the $A\to WW$ and $A\to ZZ$
branching ratios as functions of
$t^U_{L-R}$  for $\tan\beta =0.2$ and different values of
$t^U_{L+R}$. Note that 
 for small values of $\tan\beta$ the couplings between $A$
and the $D$-type vector-quarks are suppressed. As a result, only the diagrams
with $U$-type vector-quarks are relevant for the $A\to ZZ$ decay in
this case and the partial
width only depends on the $U$-type $t$ parameters $t^U_{L+R}$ and
$t^U_{L-R}$. The biggest contribution to the $A\to ZZ$ partial width
is, however, due to the top-quark loop, while the contributions
from vector-quark loops are one order of magnitude smaller. Consequently the
value of the branching ratio is almost the same as the one in
Sect.~\ref{sec:res:THDM} and the dependence on the $t$ parameters is very weak. The
smallness of the vector-quark corrections to the $A\to ZZ$ decay is mainly due
to the relation between the Yukawa couplings and the mixing angles: whenever a
Yukawa coupling gets large, some other mixing-angle dependent factor in the
couplings gets small.

For the $WW$ final state the diagrams contain both $U$ and $D$-type
vector-quarks and therefore depend on the $U$-type mixing angles, too.
  For large
Yukawa couplings the vector-quark contributions to the $A\to WW$ decay are then
dominated by just a few diagrams, where the suppression by mixing angles can be
avoided altogether. As a result they can be much larger than the corrections to
the $A\to ZZ$ decay. For small $\tan\beta$ they can be of the same order as the
contributions from SM fermion loops. 
As Fig.~\ref{fig:res:tUD-thdmvq-smalltb} shows,
the branching ratio can then reach up to about
$2\%$, i.e.\ more than twice its corresponding value in the 2HDM of
Sect.~\ref{sec:res:THDM}. For
both the $ZZ$ and $WW$ final state  we observe that the vector-quark loop
contributions can partially cancel the contributions from SM fermion loops for
$t^U_{L+R}<0$, while for $t^U_{L+R}>0$ they always add.

\begin{figure}
  \centering
  \includegraphics{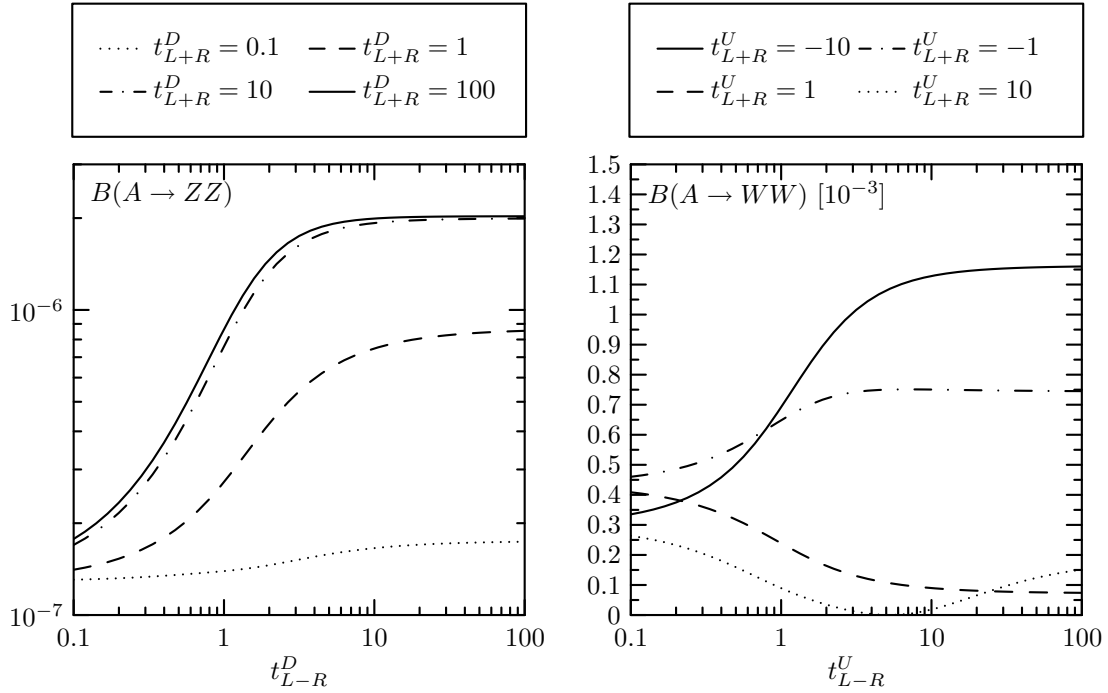}
  \caption{Results for $A\to WW,ZZ$ branching ratios in the 2HDM with
    vector-quarks for  $m_A=250$ GeV, $\tan\beta=5$ and the parameters from \eqref{eq:res:thdmvq-alpha-MQ}
    and \eqref{eq:res:thdmvq-lhparam}. Left: the
    $A\to ZZ$ branching ratio for different values of $t^D_{L+R}$ as functions
    of $t^D_{L-R}$. Right: the $A\to WW$ branching ratio for different values of
    $t^U_{L+R}$ as functions of $t^U_{L-R}$.}
  \label{fig:res:tUD-thdmvq-largetb}
\end{figure}

As  discussed in Sect.~\ref{sec:res:THDM}, the contributions of SM fermion
loops to the $A\to WW,ZZ$ partial widths drop by several 
 orders of magnitude if
we increase $\tan\beta$. Like in the 2HDM with a 4th generation of chiral
fermions, the contributions of the vector-quark loops get large for small and
large values of $\tan\beta$ if the $U$ and $D$-type vector-quark masses are of
the same order. Consequently, for large $\tan\beta$, the dominant contributions
to the $A\to WW,ZZ$ partial widths come from the vector-quark loops. 
Fig.~\ref{fig:res:tUD-thdmvq-largetb} shows 
 the $A\to WW,ZZ$ branching ratios for
$\tan\beta=5$.  Here the contributions from diagrams where the
$A$ couples to $U$-type vector-quarks are negligible and consequently the $A\to
ZZ$ branching ratio is insensitive to the $U$-type $t$ parameters $t^U_{L+R}$
and $t^U_{L-R}$. The left plot in Fig.~\ref{fig:res:tUD-thdmvq-largetb} shows
the branching ratio for $A\to ZZ$ as a function of $t^D_{L-R}$ for different
values of $t^D_{L+R}$. It  does not exceed $2 \times 10^{-6}$.

The right plot in Fig.~\ref{fig:res:tUD-thdmvq-largetb} shows the branching
ratio for $A\to WW$ as a function of $t^U_{L-R}$ for different values of
$t^U_{L+R}$. The branching ratio gets largest for $t^U_{L+R}<-10$ and
$t^U_{L+R},t^U_{L-R}>10$ and reaches approximately $0.12\%$. The dominant
diagram for $t^U_{L+R}<0$ is diagram (b) from Fig.~\ref{fig:res:vqWW} with
$(i,j,k)=(1,1,2)$. For $t^U_{L+R}>0$ it is diagram (b) with $i=j=k=1,2$. By
varying $t^D_{L+R}$ and $t^D_{L-R}$ the scale of the curves changes while their
shape stays essentially the same. However, the resulting values of the
 branching ratios
  do not exceed  $\sim 0.12\%$. As for the dependence on particle
masses we observe that the $A\to WW$ partial width is virtually independent of
the lightest vector-quark masses $m_{U_2}$ and $m_{D_2}$, while the partial
width for the $ZZ$ final state increases by one order of magnitude if we set
$m_{D_2}=\unit{200}{GeV}$. Again, the shapes of the curves in the left plot of
Fig.~\ref{fig:res:tUD-thdmvq-largetb} stay the same if we change $m_{D_2}$.

\begin{figure}
  \centering
  \includegraphics{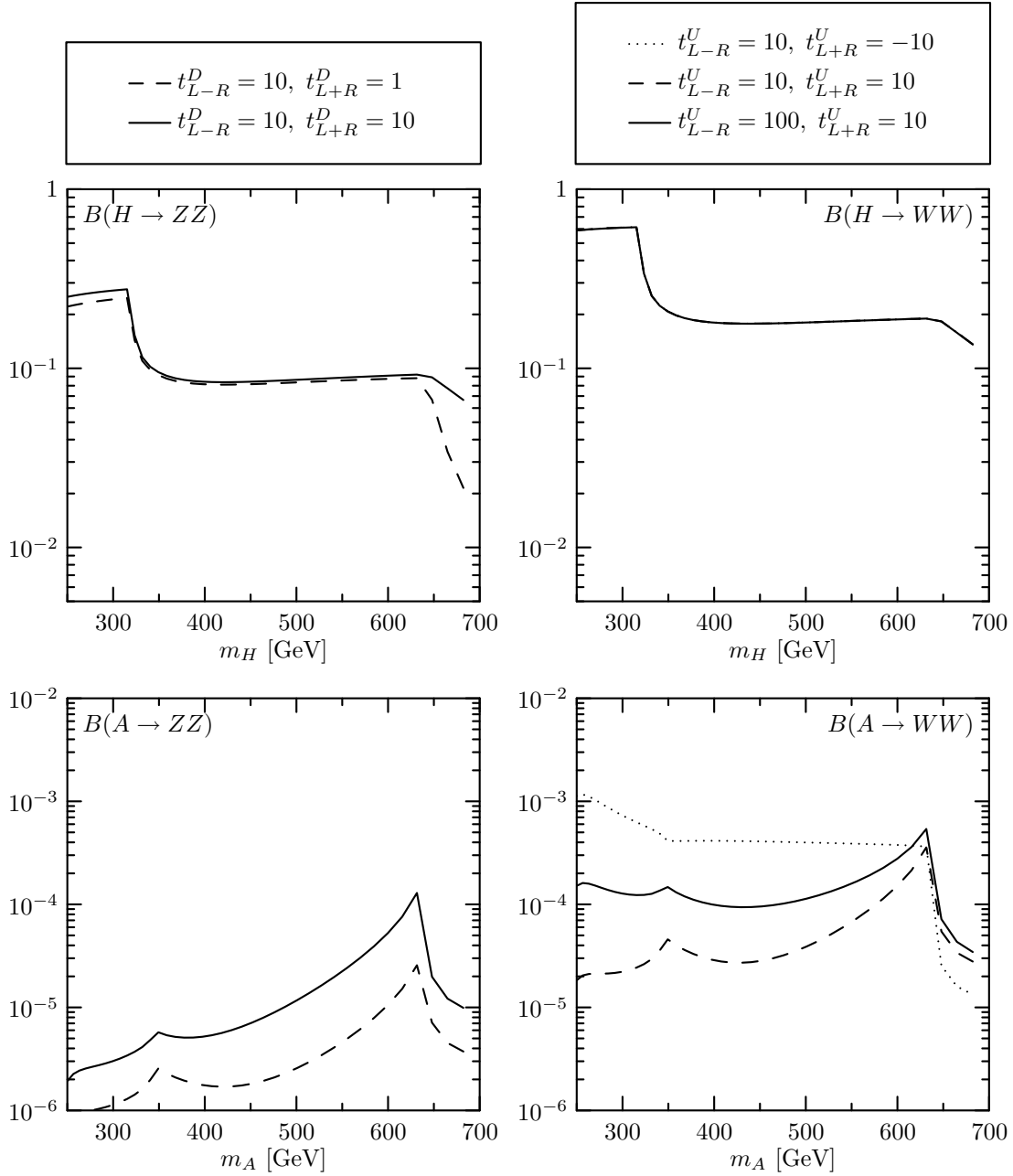}
  \caption{The $H$ (first row) and $A$ (second row) branching ratios
    in the 2HDM
    with vector-quarks for $\tan\beta=5$ and the parameters from
    \eqref{eq:res:thdmvq-alpha-MQ} and \eqref{eq:res:thdmvq-lhparam} as
    functions of the mass of the decaying particle. Left column: the branching
    ratios for the $ZZ$ final state for different combinations of $t^D_{L+R}$
    and $t^D_{L-R}$. Right column: the branching ratio for the $WW$ final state
    for different combinations of $t^U_{L+R}$ and $t^U_{L-R}$. In the plot for
    $B(H\to WW)$ all three graphs lie on top of each other.}
  \label{fig:res:mA0-thdmvq}
\end{figure}

Let us now have a look at the dependence of the branching ratios on the mass of
the decaying particles. We will only discuss the case $\tan\beta=5$ where the
$A\to WW,ZZ$ partial widths are dominated by vector-quark loops. For small
$\tan\beta$ these contributions show a similar behaviour, and the $m_A$
dependence of the SM fermion loop contributions was discussed in 
Sect.~\ref{sec:res:THDM}.  The curves of Fig.~\ref{fig:res:tUD-thdmvq-largetb}
simply scale if we change $m_A$ while their shape stays 
 essentially the same. It
is therefore sufficient to plot the $A$ branching ratios as functions of $m_A$
for a few selected combinations of the parameters
\eqref{eq:res:thdmvq-param}. Fig.~\ref{fig:res:mA0-thdmvq} shows the $A$ and
$H$ branching ratios into $WW$ and $ZZ$ for the assignments 
\eqref{eq:res:thdmvq-alpha-MQ}, $m_{U_2}=m_{D_2}=\unit{320}{GeV}$, and
different
combinations of the $t$ parameters.

Sufficiently heavy $H$, $A$ will
  also decay into   the lightest $U$- and $D$-type
vector-quarks: $H, A\to U_2\bar U_2, D_2\bar D_2.$
For large $\tan\beta$  the decay
 $H\to U_2\bar U_2$ decay is strongly suppressed 
and the total width of $H$
 depends only  on the $D$-type $t$ parameters.
  As a result the three graphs for $B(H\to WW)$ are essentially identical
because $t^D_{L+R}$ and $t^D_{L-R}$ are held fixed in the right 
column of Fig.~\ref{fig:res:mA0-thdmvq}. Below the vector-quark threshold the $H\to ZZ$
branching ratio depends on the $t$ parameters only through the contributions of
vector-quark mediated $H\to gg$ decays to the total width.  Since the $H\to gg$
channel is subdominant for large $\tan\beta$,
  the $H\to ZZ$ branching ratio is
not very sensitive to the $t$ parameters below the  vector-quark
threshold. Above that threshold it drops off more quickly for small values of
$t^D_{L+R}$. Furthermore, the $H\to WW$ and $H\to ZZ$ branching ratios both drop
by one order of magnitude at $m_H=\unit{320}{GeV}$. This corresponds to the
opening of the $H\to hh$ decay channel, which is strongly enhanced by the factor
$1/\sin(2\beta)$ in the $Hhh$ vertex. As already discussed in
 Sect.~\ref{sec:res:THDM} , this enhancement is not a necessary
 feature of the model. The
  ratios $B(A\to WW, ZZ)$   drop by an order of
magnitude  for $m_A > 2 m_{Q_2}$.

\begin{figure}
  \centering
  \includegraphics{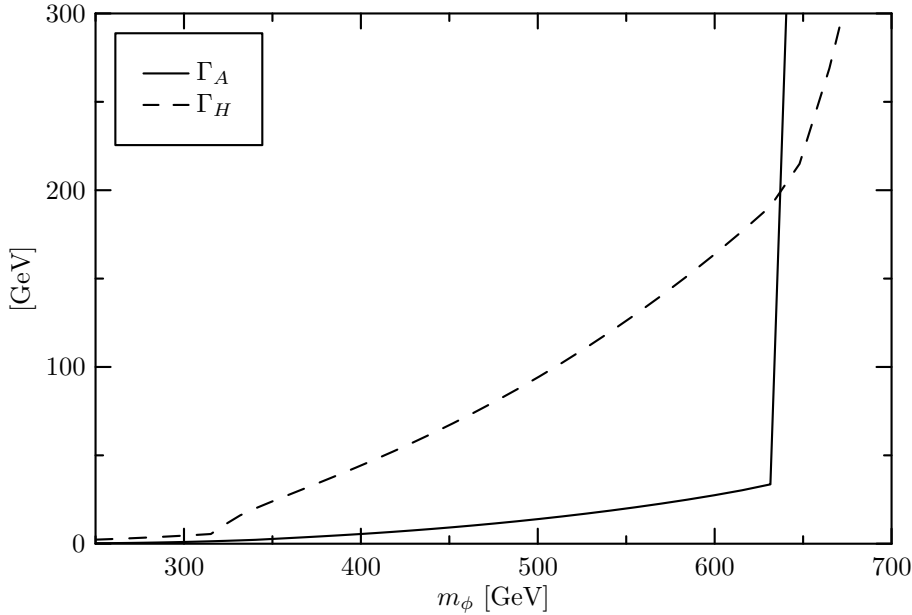}
 \caption{The total widths of $H$ and $A$ 
    in the 2HDM with vector-quarks for $\tan\beta=5$ and the parameters from
    \eqref{eq:res:thdmvq-alpha-MQ} and \eqref{eq:res:thdmvq-lhparam} as
    functions of the particle mass.}
 \label{fig:gamAH-thdmvq}
\end{figure}

Fig.~\ref{fig:gamAH-thdmvq} shows the  total widths of $H$ and $A$ 
 as a function of $m_{H,A}$ for $\tan\beta=5$. As we have allowed for 
 an enhanced $Hhh$ coupling, the width $\Gamma_H$ increases rapidly
  for $m_H > 2 m_h$. The width $\Gamma_A$ of the pseudoscalar 
   $\Gamma_A \lesssim 20$ GeV  below the vector-quark production 
  threshold which we put
  $2m_{Q_2}=640$ GeV. Above this threshold and for
 strong $AQ_2{\bar Q_2}$ coupling,  $\Gamma_A$ increases dramatically
 and the perturbative result displayed in Fig.~\ref{fig:gamAH-thdmvq}
  becomes unreliable. Yet, we emphasize that  a significantly higher $Q_2 {\bar Q}_2$
  threshold would not change our results on $B(H,A\to VV)$ 
   below $ m_{H,A}\leq 2m_{Q_2}$ in an essential way.

In summary,  for $\tan\beta=5$  we find  $B(A\to WW) \lesssim 10^{-3}$ and
 $B(A\to ZZ) \lesssim 10^{-4}$. As shown above $B(A\to WW)$  can
 become an order of magnitude larger for 
   $\tan\beta \sim 0.2$. The ratio $R_W$ can become $\sim 10\%$
if the $H\to hh$ channel is open and enhanced but the decay $A\to t\bar t$ is kinematically
forbidden.

%%%%%%%%%%%%%%%%%%%%%%%%%%%%%%%%%%%%%%%%%%%%%%%%%%%%%%%%%%%%%%%%%%%%%%%%%%%%%%%%
\subsection{Top-color assisted technicolor}
%%%%%%%%%%%%%%%%%%%%%%%%%%%%%%%%%%%%%%%%%%%%%%%%%%%%%%%%%%%%%%%%%%%%%%%%%%%%%%%%
%
An alternative to the Higgs mechanism is 
EWSB triggered by the condensation of (new) fermion antifermion pairs.
Phenomenologically viable models of this type
 include models based on the concept of top-color assisted technicolor (TC2)
\cite{Hill:1994hp,Hill:2002ap}. These models
  have two strongly interacting sectors in order to
explain EWSB and the large top-quark mass. Technicolor interactions (TC) are
responsible, via the condensation of techni-fermions, $\langle {\bar T}
T\rangle$ ($T = U, D)$, for most of EWSB, but they contribute very little to the
top-quark mass $m_t$, while top-color interactions generate through condensation
of top-quark pairs, $\langle {\bar t} t \rangle$, the bulk of $m_t$ but make
only a small contribution to EWSB.  The spin-zero states of the model are
bound-states of the techni-fermions and of $t, b$. These two sets of
bound-states form two $SU(2)_L$ doublets $\Phi_{TC}, \Phi _t$, whose couplings
to the weak gauge bosons and to $t$ and $b$ are formally equivalent to
those of a
two-Higgs doublet model. The physical spin-zero states are i) a heavy neutral
scalar $H_{TC}$ with a mass of order 1 TeV, ii) a neutral scalar $H_t$ which is
a ${\bar t} t$ bound state whose mass is expected to be of the order $m_{H_t}
\sim 2 m_t$ when estimated \`a la Nambu-Jona-Lasinio, but could in fact be
lighter \cite{Chivukula:1998wd}, and iii) a neutral ``top-pion'' $\Pi^0$ and a
pair of charged ones, $\Pi^\pm$, whose masses are predicted to lie in the range
of 200 - 300 GeV \cite{Hill:1994hp,Buchalla:1995dp}.
%  Below we   shall use $m_{H_t} \geq 120$ GeV,
%  $m_{\Pi^0}= m_{\Pi^+} \geq 180$ GeV, and $m_{H_{TC}} = 1$ TeV.

The couplings of spin-zero states to the weak gauge bosons and to the $t$ and
$b$ quarks can be obtained from an effective ${\rm SU(2)_L \ \times U(1)_Y}$
invariant Lagrangian involving the doublets $\Phi_{TC}, \Phi _t$
\cite{Leibovich:2001ev}.  The interactions of the top quark with $H_t$ and $\Pi^0$
are given by:
\begin{equation} \label{yuk-tc2}
  \Lcal_{Y,t} = 
  - \frac{Y_t}{\sqrt 2} {\bar t} t \,  H_t
  -  \frac{Y_\pi}{\sqrt 2}  {\bar t}i\gamma_5 t \,\Pi^0
  \eqpunct,
\end{equation}
where $Y_\pi =(Y_t v_T -\epsilon_t f_\pi)/v$ and $(Y_t f_\pi + \epsilon_t
v_T)/\sqrt{2} = m_t.$ Here $f_\pi$ denotes the value of the top-quark condensate
which is estimated in the TC2 model to lie between $40 \, {\rm GeV} \lesssim f_\pi
\lesssim 80 \, {\rm GeV}$ \cite{Hill:1994hp,Leibovich:2001ev}. Once $f_\pi$ is
fixed, $v_T$ is determined by the EWSB requirement that $f_\pi^2 + v_T^2 = v^2 =
(246 \, {\rm GeV})^2$. The parameter $\epsilon_t$ denotes the technicolor
contribution to the top mass which is small, by construction of the TC2 model,
and we may henceforth safely put $\epsilon_t=0$. The large top-quark mass thus
amounts to large top Yukawa couplings $Y_t$, $Y_\pi$, e.g., $Y_t\simeq
Y_\pi\simeq 4$ for $f_\pi\simeq$ 60 GeV.  The couplings of $H_t$ and $\Pi^0$ to
$b$ quarks are, on the other hand, significantly suppressed as compared with the
SM Higgs $b\bar b$ coupling. By construction, the top-color interactions do not
generate a direct contribution to the mass of the $b$ quark.
  The bulk of $m_b$ is assumed to be due to technicolor
interactions. Disregarding top-color instanton effects, the tree-level $H_tb
\bar b$ coupling is in fact zero, and the coupling of $\Pi^0$ to $b$ quarks is
given by
\begin{equation} \label{yukb-tc2}
  \Lcal_{Y,b} = - \epsilon_b \frac{f_\pi}{\sqrt{2}v} {\bar b}i
    \gamma_5 b \, \Pi^0
  \eqpunct,
\end{equation}
where $\epsilon_b = m_b \sqrt{2}/v_T$. With $m_b=\unit{4.8}{GeV}$ and
$f_\pi\leq\unit{80}{GeV}$ one gets $\epsilon_b \leq 0.03$.

In the following we consider top-pions $\Pi^{0,\pm}$ with masses
$m_{\Pi^0}=m_{\Pi^+}$ in the range between 200 GeV and 400 GeV and choose, for
reasons of comparison, the mass of $H_t$ to lie in the same range. As the
``technicolor Higgs'' boson $H_{TC}$ is much heavier than $\Pi^{0,\pm}$ and
$H_t$, it plays no role in the decays of $\Pi^0$ and $H_t$.

While the decays $H_t \to WW, ZZ$ occur already at Born level with couplings
$h_{WW}= h_{ZZ}= f_\pi/v$ (cf. Eq.~\ref{eq:gen:Hwidth}), the corresponding
decays of $\Pi^0$ are loop-induced and the amplitudes are completely dominated
by the strong $\Pi^0 {\bar t} t$ coupling, i.e., by the diagrams that correspond
to Fig.~\ref{fig:dgmWW} (b) and \ref{fig:dgmZZ} (b), respectively. This is also
the case for the $H_t, \Pi^0 \to gg$ decays. As emphasized above the decays
$\Pi^0 \to b \bar b$ are strongly suppressed and $H_t \to b \bar b$ is absent at
tree-level.

For $m_{\Pi}\leq 2 m_t$ the total rate $\Gamma_{\Pi^0}$, given in
Table~\ref{tab:BHTC.tc2}, is dominated by $\Pi^0 \to gg$.  Below the $t\bar t$
threshold the channel $H_t \to gg$ contributes also a significant portion to
$\Gamma_{H_t}$.  If $2 m_V < m_{H_t}\leq 2 m_t$ then $B(H_t \to gg) \leq 45\%$
for $f_\pi \geq 40$ GeV.

\begin{table}
  \centering
  \begin{tabular}{lMrMr}
    \hlx{hh}
      $m_{\phi}$ [GeV]:       &    200 &   300 \\
    \hlx{h}
      $\Gamma_{H_t}$ [GeV]:   &  0.072 & 0.375 \\
      $\Gamma_{\Pi^0}$ [GeV]: &  0.088 & 0.500 \\
    \hlx{hh}
  \end{tabular}
  \caption{The total widths of $\phi =H_t, \Pi^0$ for two values of $m_\phi$ and
    $f_\pi=\unit{40}{GeV}$.}
  \label{tab:BHTC.tc2}
\end{table}

\begin{figure}
  \centering
  \includegraphics{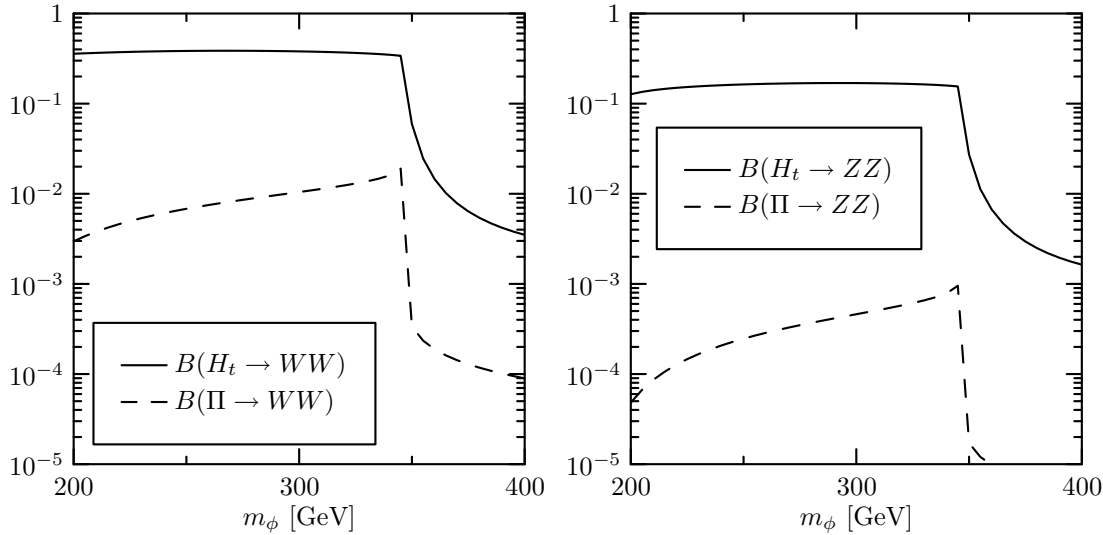}
  \caption{The TC2 branching ratios of the decays of $\Pi^0$ and $H_{t}$ into
    $WW$ (left) and $ZZ$ (right) final states as functions of the respective
    mass ($\phi= \Pi^0, H_t$), for $f_\pi=\unit{40}{GeV}$.}
  \label{fig:mpi.tc2}
\end{figure}

As a consequence of these features, the branching ratios $B(\Pi^0 \to WW,ZZ)$
are essentially independent of $f_\pi$, while $B(H_t \to WW,ZZ)$ increase with
increasing $f_\pi$, namely by about a factor of two if $f_\pi$ is changed from
\unit{40}{GeV} to \unit{80}{GeV}.

Fig.~\ref{fig:mpi.tc2} shows the branching ratios $B(H_t, \Pi^0 \to WW/ZZ)$ as
functions of the mass of the respective spin zero resonance, for $f_\pi =
\unit{40}{GeV}$. Once the $t \bar t$ threshold is crossed, there are sharp
drops, because the decays into $t \bar t$ then dominate the total widths of both
${\Pi^0}$ and $H_t$.  For masses below the $t\bar t$ threshold, which are
preferred by model predictions \cite{Hill:1994hp,Buchalla:1995dp}, we thus obtain
$B(\Pi^0 \to WW) \lesssim 2 \times 10^{-2}$, $B(\Pi^0 \to ZZ) \lesssim 10^{-3}$
and the ratios $R_W \lesssim 0.07$, $R_Z \lesssim 0.007$.
%

%
%
%%%%%%%%%%%%%%%%%%%%%%%%%%%%%%%%%%%%%%%%%%%%%
\section{Summary and Conclusions} \label{sec:conclusions}
%%%%%%%%%%%%%%%%%%%%%%%%%%%%%%%%%%%%%%%%%%%%%%

In this paper we addressed two questions: Assuming the existence of a
non-standard EWSB sector with a spin-0 (Higgs) resonance
spectrum which includes a pseudoscalar $A$ with mass $m_A \gtrsim
200$ GeV, we investigated 
  whether the decays  $A\to WW/ZZ$ can have branching ratios at the
 percent level and analysed 
the size of $B(A \to WW/ZZ)$ relative to the respective branching
ratio of a scalar boson $H$ with mass $m_H \simeq m_A$. 
 We analyzed this in
the context of the MSSM, in non-supersymmetric type-II two-Higgs
doublet extensions of the SM without and with additional heavy
fermions, and in TC2.

For the MSSM we confirmed what is known in the literature, i.e., that
 the  $A\to WW/ZZ$ are rare. But also the branching ratios
  $B(H \to WW/ZZ)$  can be below the per mill level in the
  scenario  discussed  in Sect.~\ref{subsec:MSSM}. In the
  three-generation 2HDM and in the corresponding models with an additional 4th
  sequential fermion generation or heavy vector-like quarks the
  branching ratio  $B(A \to WW)$ can reach about $2\%$ for $m_A \leq 2
  m_t$, while $B(A \to ZZ)\lesssim 10^{-3}$. In this case the total
  width $\Gamma_A$ is dominated by $A\to gg$ and $A\to Z h$
  (if kinematically possible) while $A \to b {\bar b}, \tau^-\tau^+$
  are suppressed. The tree-level couplings $HVV$
 can be severely suppressed, which implies small 
   $B(H \to WW/ZZ)$ and ratios $R_{W,Z}$
 (\ref{rvratio}) of order one -- but this amounts to fine-tuned
 parameters $\alpha$ and $\beta$. Even if the $HVV$ Born couplings are
 not suppressed,  $B(H \to WW)$ and $B(H \to ZZ)$ may  not exceed
  $\sim 10\%$ below the $t\bar t$ threshold 
   if $H\to hh$ is kinematically possible and enhanced. 
  In TC2 models the latter decay mode is absent. If a composite
  scalar resonance $H_t$ below the $t \bar t$ threshold exists, it
  will predominantly decay into $gg, WW, ZZ$. We obtain
  $B(H_t\to WW) \lesssim 0.4$ and  $B(H_t\to ZZ) \lesssim 0.1$.
  The neutral pseudoscalar resonance $\Pi^0$ with $m_\Pi \leq 2m_t$ decays dominantly into
  gluons;  $B(\Pi^0\to WW) \lesssim 0.02$ and 
    $B(\Pi^0\to ZZ) \lesssim 10^{-3}$.

In view of these results one
  may resume the issue raised in the introduction, whether the discovery
of a spin-zero resonance $\phi$ in the $WW$ and/or $ZZ$ channel allows the
conclusion that this particle has  
$J^{PC}=0^{++}$  -- prior to performing the tests proposed in  \cite{Nelson:1986ki,
  Soni:1993jc, Skjold:1993jd, Barger:1993wt, Arens:1994wd, Choi:2002jk,
  Buszello:2002uu, Godbole:2007cn, Accomando:2006ga}. Our findings
suggest the following: If $\phi \to WW, ZZ$ are dominant decay channels then
it is very likely that $\phi$ is a scalar. If these modes are
suppressed as compared to other channels, there is a non-negligible
probability that $\phi$ is a pseudoscalar. 
Of course, these CP tests -- and other CP tests in 
 $\phi \to t\bar t, \tau^-\tau^+$ 
  (see .e.g., \cite{Accomando:2006ga,Berge:2008dr,Berge:2008wi}) --
are eventually indispensable in order to establish whether or not
 $\phi$ is a true CP eigenstate.

%
%
%%%%%%%%%%%%%%%%%%%%%%%%%%%%%%%%%%%%%
\subsubsection*{Acknowledgments}
%%%%%%%%%%%%%%%%%%%%%%%%%%%%%%%%%%%
This work was supported by Deutsche Forschungsgemeinschaft 
 SFB/TR9.

%
%%%%%%%%%%%%%%%%%%%%%%%%%%%%%%%%%%%%%%%%%%%%%%%%%%%%%%%
\begin{appendix}
\section{Feynman Rules for the Vector-Quark Model}
\label{sec:thdmvq-rules}
In this appendix we list the Feynman rules for the 2HDM model with vector-like
quarks, which we discussed in section \ref{sec:res:thdmvq}. We list
  only the couplings of vector-quarks to $H$, $A$, $W$
and $Z$. In the rules given, all particles flow into the vertex. In the
case where two vertices are related by a hermitian conjugation, only one
representative is shown. As usual, $P_{R,L}=(1\pm\gamma_5)/2$.

\input{THDMVQ.tex}

\end{appendix}

%\nocite{Djouadi05}
\par
\bibliography{Adecay}
\end{document}

%% file: dgmWW.tex
\unitlength=1bp%

\begin{feynartspicture}(400,80)(5,1)

\FADiagram{(a)}
\FAProp(0.,10.)(6.5,10.)(0.,){/ScalarDash}{0}
\FALabel(3.25,9.18)[t]{$A$}
\FAProp(20.,15.)(13.,14.)(0.,){/Sine}{-1}
\FALabel(16.2808,15.5544)[b]{$W^+$}
\FAProp(20.,5.)(13.,6.)(0.,){/Sine}{1}
\FALabel(16.2808,4.44558)[t]{$W^-$}
\FAProp(6.5,10.)(13.,14.)(0.,){/Straight}{-1}
\FALabel(9.20801,13.1807)[br]{$\tau$}
\FAProp(6.5,10.)(13.,6.)(0.,){/Straight}{1}
\FALabel(9.20801,6.81927)[tr]{$\tau$}
\FAProp(13.,14.)(13.,6.)(0.,){/Straight}{-1}
\FALabel(14.274,10.)[l]{$\nu_\tau$}
\FAVert(6.5,10.){0}
\FAVert(13.,14.){0}
\FAVert(13.,6.){0}

\FADiagram{(b)}
\FAProp(0.,10.)(6.5,10.)(0.,){/ScalarDash}{0}
\FALabel(3.25,9.18)[t]{$A$}
\FAProp(20.,15.)(13.,14.)(0.,){/Sine}{-1}
\FALabel(16.2808,15.5544)[b]{$W^+$}
\FAProp(20.,5.)(13.,6.)(0.,){/Sine}{1}
\FALabel(16.2808,4.44558)[t]{$W^-$}
\FAProp(6.5,10.)(13.,14.)(0.,){/Straight}{1}
\FALabel(9.20801,13.1807)[br]{$t$}
\FAProp(6.5,10.)(13.,6.)(0.,){/Straight}{-1}
\FALabel(9.20801,6.81927)[tr]{$t$}
\FAProp(13.,14.)(13.,6.)(0.,){/Straight}{1}
\FALabel(14.274,10.)[l]{$b$}
\FAVert(6.5,10.){0}
\FAVert(13.,14.){0}
\FAVert(13.,6.){0}

\FADiagram{(c)}
\FAProp(0.,10.)(6.5,10.)(0.,){/ScalarDash}{0}
\FALabel(3.25,9.18)[t]{$A$}
\FAProp(20.,15.)(13.,14.)(0.,){/Sine}{-1}
\FALabel(16.2808,15.5544)[b]{$W^+$}
\FAProp(20.,5.)(13.,6.)(0.,){/Sine}{1}
\FALabel(16.2808,4.44558)[t]{$W^-$}
\FAProp(6.5,10.)(13.,14.)(0.,){/Straight}{-1}
\FALabel(9.20801,13.1807)[br]{$b$}
\FAProp(6.5,10.)(13.,6.)(0.,){/Straight}{1}
\FALabel(9.20801,6.81927)[tr]{$b$}
\FAProp(13.,14.)(13.,6.)(0.,){/Straight}{-1}
\FALabel(14.274,10.)[l]{$t$}
\FAVert(6.5,10.){0}
\FAVert(13.,14.){0}
\FAVert(13.,6.){0}

\FADiagram{(d)}
\FAProp(0.,10.)(6.5,10.)(0.,){/ScalarDash}{0}
\FALabel(3.25,9.18)[t]{$A$}
\FAProp(20.,15.)(13.,14.)(0.,){/Sine}{-1}
\FALabel(16.2808,15.5544)[b]{$W^+$}
\FAProp(20.,5.)(13.,6.)(0.,){/Sine}{1}
\FALabel(16.2808,4.44558)[t]{$W^-$}
\FAProp(6.5,10.)(13.,14.)(0.,){/Straight}{0}
\FALabel(9.33903,12.9678)[br]{$\tilde \chi^0$}
\FAProp(6.5,10.)(13.,6.)(0.,){/Straight}{0}
\FALabel(9.33903,7.03218)[tr]{$\tilde \chi^0$}
\FAProp(13.,14.)(13.,6.)(0.,){/Straight}{1}
\FALabel(14.274,10.)[l]{$\tilde \chi^-$}
\FAVert(6.5,10.){0}
\FAVert(13.,14.){0}
\FAVert(13.,6.){0}

\FADiagram{(e)}
\FAProp(0.,10.)(6.5,10.)(0.,){/ScalarDash}{0}
\FALabel(3.25,9.18)[t]{$A$}
\FAProp(20.,15.)(13.,14.)(0.,){/Sine}{-1}
\FALabel(16.2808,15.5544)[b]{$W^+$}
\FAProp(20.,5.)(13.,6.)(0.,){/Sine}{1}
\FALabel(16.2808,4.44558)[t]{$W^-$}
\FAProp(6.5,10.)(13.,14.)(0.,){/Straight}{-1}
\FALabel(9.20801,13.1807)[br]{$\tilde \chi^-$}
\FAProp(6.5,10.)(13.,6.)(0.,){/Straight}{1}
\FALabel(9.20801,6.81927)[tr]{$\tilde \chi^-$}
\FAProp(13.,14.)(13.,6.)(0.,){/Straight}{0}
\FALabel(14.024,10.)[l]{$\tilde \chi^0$}
\FAVert(6.5,10.){0}
\FAVert(13.,14.){0}
\FAVert(13.,6.){0}
\end{feynartspicture}

%% file: dgmZZ.tex
\unitlength=1bp%

\begin{feynartspicture}(400,80)(5,1)

\FADiagram{(a)}
\FAProp(0.,10.)(6.5,10.)(0.,){/ScalarDash}{0}
\FALabel(3.25,9.18)[t]{$A$}
\FAProp(20.,15.)(13.,14.)(0.,){/Sine}{0}
\FALabel(16.2808,15.5544)[b]{$Z$}
\FAProp(20.,5.)(13.,6.)(0.,){/Sine}{0}
\FALabel(16.2808,4.44558)[t]{$Z$}
\FAProp(6.5,10.)(13.,14.)(0.,){/Straight}{1}
\FALabel(9.20801,13.1807)[br]{$\tau$}
\FAProp(6.5,10.)(13.,6.)(0.,){/Straight}{-1}
\FALabel(9.20801,6.81927)[tr]{$\tau$}
\FAProp(13.,14.)(13.,6.)(0.,){/Straight}{1}
\FALabel(14.274,10.)[l]{$\tau$}
\FAVert(6.5,10.){0}
\FAVert(13.,14.){0}
\FAVert(13.,6.){0}

\FADiagram{(b)}
\FAProp(0.,10.)(6.5,10.)(0.,){/ScalarDash}{0}
\FALabel(3.25,9.18)[t]{$A$}
\FAProp(20.,15.)(13.,14.)(0.,){/Sine}{0}
\FALabel(16.2808,15.5544)[b]{$Z$}
\FAProp(20.,5.)(13.,6.)(0.,){/Sine}{0}
\FALabel(16.2808,4.44558)[t]{$Z$}
\FAProp(6.5,10.)(13.,14.)(0.,){/Straight}{1}
\FALabel(9.20801,13.1807)[br]{$t$}
\FAProp(6.5,10.)(13.,6.)(0.,){/Straight}{-1}
\FALabel(9.20801,6.81927)[tr]{$t$}
\FAProp(13.,14.)(13.,6.)(0.,){/Straight}{1}
\FALabel(14.274,10.)[l]{$t$}
\FAVert(6.5,10.){0}
\FAVert(13.,14.){0}
\FAVert(13.,6.){0}

\FADiagram{(c)}
\FAProp(0.,10.)(6.5,10.)(0.,){/ScalarDash}{0}
\FALabel(3.25,9.18)[t]{$A$}
\FAProp(20.,15.)(13.,14.)(0.,){/Sine}{0}
\FALabel(16.2808,15.5544)[b]{$Z$}
\FAProp(20.,5.)(13.,6.)(0.,){/Sine}{0}
\FALabel(16.2808,4.44558)[t]{$Z$}
\FAProp(6.5,10.)(13.,14.)(0.,){/Straight}{1}
\FALabel(9.20801,13.1807)[br]{$b$}
\FAProp(6.5,10.)(13.,6.)(0.,){/Straight}{-1}
\FALabel(9.20801,6.81927)[tr]{$b$}
\FAProp(13.,14.)(13.,6.)(0.,){/Straight}{1}
\FALabel(14.274,10.)[l]{$b$}
\FAVert(6.5,10.){0}
\FAVert(13.,14.){0}
\FAVert(13.,6.){0}

\FADiagram{(d)}
\FAProp(0.,10.)(6.5,10.)(0.,){/ScalarDash}{0}
\FALabel(3.25,9.18)[t]{$A$}
\FAProp(20.,15.)(13.,14.)(0.,){/Sine}{0}
\FALabel(16.2808,15.5544)[b]{$Z$}
\FAProp(20.,5.)(13.,6.)(0.,){/Sine}{0}
\FALabel(16.2808,4.44558)[t]{$Z$}
\FAProp(6.5,10.)(13.,14.)(0.,){/Straight}{0}
\FALabel(9.33903,12.9678)[br]{$\tilde \chi^0$}
\FAProp(6.5,10.)(13.,6.)(0.,){/Straight}{0}
\FALabel(9.33903,7.03218)[tr]{$\tilde \chi^0$}
\FAProp(13.,14.)(13.,6.)(0.,){/Straight}{0}
\FALabel(14.024,10.)[l]{$\tilde \chi^0$}
\FAVert(6.5,10.){0}
\FAVert(13.,14.){0}
\FAVert(13.,6.){0}

\FADiagram{(e)}
\FAProp(0.,10.)(6.5,10.)(0.,){/ScalarDash}{0}
\FALabel(3.25,9.18)[t]{$A$}
\FAProp(20.,15.)(13.,14.)(0.,){/Sine}{0}
\FALabel(16.2808,15.5544)[b]{$Z$}
\FAProp(20.,5.)(13.,6.)(0.,){/Sine}{0}
\FALabel(16.2808,4.44558)[t]{$Z$}
\FAProp(6.5,10.)(13.,14.)(0.,){/Straight}{1}
\FALabel(9.20801,13.1807)[br]{$\tilde \chi^-$}
\FAProp(6.5,10.)(13.,6.)(0.,){/Straight}{-1}
\FALabel(9.20801,6.81927)[tr]{$\tilde \chi^-$}
\FAProp(13.,14.)(13.,6.)(0.,){/Straight}{1}
\FALabel(14.274,10.)[l]{$\tilde \chi^-$}
\FAVert(6.5,10.){0}
\FAVert(13.,14.){0}
\FAVert(13.,6.){0}
\end{feynartspicture}

%% file: vqWW.tex
\unitlength=1bp%

\begin{feynartspicture}(160,80)(2,1)

\FADiagram{(a)}
\FAProp(0.,10.)(6.5,10.)(0.,){/ScalarDash}{0}
\FALabel(3.25,9.18)[t]{$A$}
\FAProp(20.,15.)(13.,14.)(0.,){/Sine}{-1}
\FALabel(16.2808,15.5544)[b]{$W^+$}
\FAProp(20.,5.)(13.,6.)(0.,){/Sine}{1}
\FALabel(16.2808,4.44558)[t]{$W^-$}
\FAProp(6.5,10.)(13.,14.)(0.,){/Straight}{1}
\FALabel(9.20801,13.1807)[br]{$U_i$}
\FAProp(6.5,10.)(13.,6.)(0.,){/Straight}{-1}
\FALabel(9.20801,6.81927)[tr]{$U_j$}
\FAProp(13.,14.)(13.,6.)(0.,){/Straight}{1}
\FALabel(14.274,10.)[l]{$D_k$}
\FAVert(6.5,10.){0}
\FAVert(13.,14.){0}
\FAVert(13.,6.){0}

\FADiagram{(b)}
\FAProp(0.,10.)(6.5,10.)(0.,){/ScalarDash}{0}
\FALabel(3.25,9.18)[t]{$A$}
\FAProp(20.,15.)(13.,14.)(0.,){/Sine}{-1}
\FALabel(16.2808,15.5544)[b]{$W^+$}
\FAProp(20.,5.)(13.,6.)(0.,){/Sine}{1}
\FALabel(16.2808,4.44558)[t]{$W^-$}
\FAProp(6.5,10.)(13.,14.)(0.,){/Straight}{-1}
\FALabel(9.20801,13.1807)[br]{$D_i$}
\FAProp(6.5,10.)(13.,6.)(0.,){/Straight}{1}
\FALabel(9.20801,6.81927)[tr]{$D_j$}
\FAProp(13.,14.)(13.,6.)(0.,){/Straight}{-1}
\FALabel(14.274,10.)[l]{$U_k$}
\FAVert(6.5,10.){0}
\FAVert(13.,14.){0}
\FAVert(13.,6.){0}
\end{feynartspicture}

%% file: vqZZ.tex
\unitlength=1bp%

\begin{feynartspicture}(320,80)(4,1)

\FADiagram{(a)}
\FAProp(0.,10.)(6.5,10.)(0.,){/ScalarDash}{0}
\FALabel(3.25,9.18)[t]{$A$}
\FAProp(20.,15.)(13.,14.)(0.,){/Sine}{0}
\FALabel(16.2808,15.5544)[b]{$Z$}
\FAProp(20.,5.)(13.,6.)(0.,){/Sine}{0}
\FALabel(16.2808,4.44558)[t]{$Z$}
\FAProp(6.5,10.)(13.,14.)(0.,){/Straight}{1}
\FALabel(9.20801,13.1807)[br]{$U_i$}
\FAProp(6.5,10.)(13.,6.)(0.,){/Straight}{-1}
\FALabel(9.20801,6.81927)[tr]{$U_j$}
\FAProp(13.,14.)(13.,6.)(0.,){/Straight}{1}
\FALabel(14.274,10.)[l]{$U_k$}
\FAVert(6.5,10.){0}
\FAVert(13.,14.){0}
\FAVert(13.,6.){0}

\FADiagram{(b)}
\FAProp(0.,10.)(6.5,10.)(0.,){/ScalarDash}{0}
\FALabel(3.25,9.18)[t]{$A$}
\FAProp(20.,15.)(13.,14.)(0.,){/Sine}{0}
\FALabel(16.2808,15.5544)[b]{$Z$}
\FAProp(20.,5.)(13.,6.)(0.,){/Sine}{0}
\FALabel(16.2808,4.44558)[t]{$Z$}
\FAProp(6.5,10.)(13.,14.)(0.,){/Straight}{-1}
\FALabel(9.20801,13.1807)[br]{$U_i$}
\FAProp(6.5,10.)(13.,6.)(0.,){/Straight}{1}
\FALabel(9.20801,6.81927)[tr]{$U_j$}
\FAProp(13.,14.)(13.,6.)(0.,){/Straight}{-1}
\FALabel(14.274,10.)[l]{$U_k$}
\FAVert(6.5,10.){0}
\FAVert(13.,14.){0}
\FAVert(13.,6.){0}

\FADiagram{(c)}
\FAProp(0.,10.)(6.5,10.)(0.,){/ScalarDash}{0}
\FALabel(3.25,9.18)[t]{$A$}
\FAProp(20.,15.)(13.,14.)(0.,){/Sine}{0}
\FALabel(16.2808,15.5544)[b]{$Z$}
\FAProp(20.,5.)(13.,6.)(0.,){/Sine}{0}
\FALabel(16.2808,4.44558)[t]{$Z$}
\FAProp(6.5,10.)(13.,14.)(0.,){/Straight}{1}
\FALabel(9.20801,13.1807)[br]{$D_i$}
\FAProp(6.5,10.)(13.,6.)(0.,){/Straight}{-1}
\FALabel(9.20801,6.81927)[tr]{$D_j$}
\FAProp(13.,14.)(13.,6.)(0.,){/Straight}{1}
\FALabel(14.274,10.)[l]{$D_k$}
\FAVert(6.5,10.){0}
\FAVert(13.,14.){0}
\FAVert(13.,6.){0}

\FADiagram{(d)}
\FAProp(0.,10.)(6.5,10.)(0.,){/ScalarDash}{0}
\FALabel(3.25,9.18)[t]{$A$}
\FAProp(20.,15.)(13.,14.)(0.,){/Sine}{0}
\FALabel(16.2808,15.5544)[b]{$Z$}
\FAProp(20.,5.)(13.,6.)(0.,){/Sine}{0}
\FALabel(16.2808,4.44558)[t]{$Z$}
\FAProp(6.5,10.)(13.,14.)(0.,){/Straight}{-1}
\FALabel(9.20801,13.1807)[br]{$D_i$}
\FAProp(6.5,10.)(13.,6.)(0.,){/Straight}{1}
\FALabel(9.20801,6.81927)[tr]{$D_j$}
\FAProp(13.,14.)(13.,6.)(0.,){/Straight}{-1}
\FALabel(14.274,10.)[l]{$D_k$}
\FAVert(6.5,10.){0}
\FAVert(13.,14.){0}
\FAVert(13.,6.){0}
\end{feynartspicture}

%% file: THDMVQ.tex
\subsection{Yukawa Couplings of $H$ to Vector-Quarks}

\begin{feynrules}
\feynrulesvertexiii{H}{\bar U_2}{U_2}{
  \frac{i s_{\alpha }}{\sqrt{2}} \bigl(c_R^U s_L^U y_U+c_L^U s_R^U \tilde{y}_U\bigr)
}%
%% \feynrulesvertexiii{H}{\bar U_2}{U_1}{
%%   -\frac{i s_{\alpha }}{\sqrt{2}} \bigl[\bigl(c_L^U c_R^U y_U-s_L^U s_R^U \tilde{y}_U\bigr) P_L + \bigl(-s_L^U s_R^U y_U+c_L^U c_R^U \tilde{y}_U\bigr) P_R \bigr]
%% }%
\feynrulesvertexiii{H}{\bar U_1}{U_2}{
  \frac{i s_{\alpha }}{\sqrt{2}} \bigl[\bigl(s_L^U s_R^U y_U-c_L^U c_R^U \tilde{y}_U\bigr) P_L + \bigl(-c_L^U c_R^U y_U+s_L^U s_R^U \tilde{y}_U\bigr) P_R \bigr]
}%
\feynrulesvertexiii{H}{\bar U_1}{U_1}{
  -\frac{i s_{\alpha }}{\sqrt{2}} \bigl(c_L^U s_R^U y_U+c_R^U s_L^U \tilde{y}_U\bigr)
}%
\feynrulesvertexiii{H}{\bar D_2}{D_2}{
  \frac{i c_{\alpha }}{\sqrt{2}} \bigl(c_R^D s_L^D y_D+c_L^D s_R^D \tilde{y}_D\bigr)
}%
%% \feynrulesvertexiii{H}{\bar D_2}{D_1}{
%%   -\frac{i c_{\alpha }}{\sqrt{2}} \bigl[\bigl(c_L^D c_R^D y_D-s_L^D s_R^D \tilde{y}_D\bigr) P_L +\bigl(-s_L^D s_R^D y_D+c_L^D c_R^D \tilde{y}_D\bigr) P_R \bigr]
%% }%
\feynrulesvertexiii{H}{\bar D_1}{D_2}{
  \frac{i c_{\alpha }}{\sqrt{2}} \bigl[\bigl(s_L^D s_R^D y_D-c_L^D c_R^D \tilde{y}_D\bigr) P_L +\bigl(-c_L^D c_R^D y_D+s_L^D s_R^D \tilde{y}_D\bigr) P_R \bigr]
}%
\feynrulesvertexiii{H}{\bar D_1}{D_1}{
  -\frac{i c_{\alpha }}{\sqrt{2}} \bigl(c_L^D s_R^D y_D+c_R^D s_L^D \tilde{y}_D\bigr)
}%
\end{feynrules}

\subsection{Yukawa Couplings of $A$ to Vector-Quarks}

\begin{feynrules}
\feynrulesvertexiii{A}{\bar U_2}{U_2}{
  \frac{c_{\beta }}{\sqrt{2}} \bigl(c_R^U s_L^U y_U-c_L^U s_R^U \tilde{y}_U\bigr) \gamma_5
}%
\feynrulesvertexiii{A}{\bar U_2}{U_1}{
  \frac{c_{\beta }}{\sqrt{2}} \bigl[\bigl(c_L^U c_R^U y_U+s_L^U s_R^U \tilde{y}_U\bigr) P_L +\bigl(s_L^U s_R^U y_U+c_L^U c_R^U \tilde{y}_U\bigr) P_R \bigr]
}%
%% \feynrulesvertexiii{A}{\bar U_1}{U_2}{
%%   -\frac{c_{\beta }}{\sqrt{2}} \bigl[\bigl(s_L^U s_R^U y_U+c_L^U c_R^U \tilde{y}_U\bigr) P_L + \bigl(c_L^U c_R^U y_U+s_L^U s_R^U \tilde{y}_U\bigr) P_R \bigr]
%% }%
\feynrulesvertexiii{A}{\bar U_1}{U_1}{
  \frac{c_{\beta }}{\sqrt{2}} \bigl(-c_L^U s_R^U y_U+c_R^U s_L^U \tilde{y}_U\bigr) \gamma_5
}%
\feynrulesvertexiii{A}{\bar D_2}{D_2}{
  \frac{s_{\beta }}{\sqrt{2}} \bigl(c_R^D s_L^D y_D-c_L^D s_R^D \tilde{y}_D\bigr) \gamma_5
}%
\feynrulesvertexiii{A}{\bar D_2}{D_1}{
  \frac{s_{\beta }}{\sqrt{2}} \bigl[\bigl(c_L^D c_R^D y_D+s_L^D s_R^D \tilde{y}_D\bigr) P_L + \bigl(s_L^D s_R^D y_D+c_L^D c_R^D \tilde{y}_D\bigr) P_R \bigr]
}%
%% \feynrulesvertexiii{A}{\bar D_1}{D_2}{
%%   -\frac{s_{\beta }}{\sqrt{2}} \bigl[\bigl(s_L^D s_R^D y_D+c_L^D c_R^D \tilde{y}_D\bigr) P_L + \bigl(c_L^D c_R^D y_D+s_L^D s_R^D \tilde{y}_D\bigr) P_R \bigr]
%% }%
\feynrulesvertexiii{A}{\bar D_1}{D_1}{
  \frac{s_{\beta }}{\sqrt{2}} \bigl(-c_L^D s_R^D y_D+c_R^D s_L^D \tilde{y}_D\bigr) \gamma_5
}%
\end{feynrules}

\subsection{Couplings of $W$ to Vector-Quarks}

\begin{feynrules}
\feynrulesvertexiii{W^+_\mu}{\bar U_2}{D_2}{
  -\frac{i e}{\sqrt{2} s_W} \gamma_\mu\bigl(s_L^D s_L^U P_L + s_R^D s_R^U P_R\bigr)
}%
\feynrulesvertexiii{W^+_\mu}{\bar U_2}{D_1}{
  \frac{i e}{\sqrt{2} s_W} \gamma_\mu\bigl(c_L^D s_L^U P_L + c_R^D s_R^U P_R\bigr)
}%
\feynrulesvertexiii{W^+_\mu}{\bar U_1}{D_2}{
  \frac{i e}{\sqrt{2} s_W} \gamma_\mu\bigl(c_L^U s_L^D P_L + c_R^U s_R^D P_R\bigr)
}%
\feynrulesvertexiii{W^+_\mu}{\bar U_1}{D_1}{
  -\frac{i e}{\sqrt{2} s_W} \gamma_\mu\bigl(c_L^D c_L^U P_L + c_R^D c_R^U P_R\bigr)
}%
%% \feynrulesvertexiii{W^-_\mu}{\bar D_2}{U_2}{
%%   -\frac{i e}{\sqrt{2} s_W} \gamma_\mu\bigl(s_L^D s_L^U P_L + s_R^D s_R^U P_R\bigr)
%% }%
%% \feynrulesvertexiii{W^-_\mu}{\bar D_2}{U_1}{
%%   \frac{i e}{\sqrt{2} s_W} \gamma_\mu\bigl(c_L^U s_L^D P_L + c_R^U s_R^D P_R\bigr)
%% }%
%% \feynrulesvertexiii{W^-_\mu}{\bar D_1}{U_2}{
%%   \frac{i e}{\sqrt{2} s_W} \gamma_\mu\bigl(c_L^D s_L^U P_L + c_R^D s_R^U P_R\bigr)
%% }%
%% \feynrulesvertexiii{W^-_\mu}{\bar D_1}{U_1}{
%%   -\frac{i e}{\sqrt{2} s_W} \gamma_\mu\bigl(c_L^D c_L^U P_L + c_R^D c_R^U P_R\bigr)
%% }%
\end{feynrules}

\subsection{Couplings of $Z$ to Vector-Quarks}

\begin{feynrules}
\feynrulesvertexiii{Z_\mu}{\bar U_2}{U_2}{
  \frac{i e}{c_W s_W} \gamma_\mu\bigl[\bigl(\tfrac23 s_W^2 - \tfrac12\bigl(s_L^U\bigr){}^2\bigr) P_L +\bigl(\tfrac23 s_W^2 - \tfrac12\bigl(s_R^U\bigr){}^2\bigr) P_R \bigr]
}%
\feynrulesvertexiii{Z_\mu}{\bar U_2}{U_1}{
  \frac{i e}{2 c_W s_W} \gamma_\mu\bigl(c_L^U s_L^U  P_L +c_R^U s_R^U  P_R \bigr)
}%
%% \feynrulesvertexiii{Z_\mu}{\bar U_1}{U_2}{
%%   \frac{i e}{2 c_W s_W} \gamma_\mu\bigl(c_L^U s_L^U  P_L +c_R^U s_R^U  P_R \bigr)
%% }%
\feynrulesvertexiii{Z_\mu}{\bar U_1}{U_1}{
  \frac{i e}{c_W s_W} \gamma_\mu\bigl[\bigl(\tfrac23 s_W^2 - \tfrac12 \bigl(c_L^U\bigr){}^2\bigr)  P_L + \bigl(\tfrac23 s_W^2 - \tfrac12 \bigl(c_R^U\bigr){}^2 \bigr)  P_R \bigr]
}%
\feynrulesvertexiii{Z_\mu}{\bar D_2}{D_2}{
  -\frac{i e}{c_W s_W} \gamma_\mu\bigl[\bigl(\tfrac13 s_W^2 - \tfrac12 \bigl(s_L^D\bigr){}^2 \bigr)  P_L + \bigl(\tfrac13 s_W^2 - \tfrac12 \bigl(s_R^D\bigr){}^2 \bigr)  P_R \bigr]
}%
\feynrulesvertexiii{Z_\mu}{\bar D_2}{D_1}{
  -\frac{i e}{2 c_W s_W} \gamma_\mu\bigl(c_L^D s_L^D  P_L +c_R^D s_R^D  P_R \bigr)
}%
%% \feynrulesvertexiii{Z_\mu}{\bar D_1}{D_2}{
%%   -\frac{i e}{2 c_W s_W} \gamma_\mu\bigl(c_L^D s_L^D  P_L +c_R^D s_R^D  P_R \bigr)
%% }%
\feynrulesvertexiii{Z_\mu}{\bar D_1}{D_1}{
  -\frac{i e}{c_W s_W} \gamma_\mu\bigl[\bigl(\tfrac13 s_W^2 - \tfrac12 \bigl(c_L^D\bigr){}^2 \bigr)  P_L +\bigl(\tfrac13 s_W^2 - \tfrac12 \bigl(c_R^D\bigr){}^2\bigr) P_R \bigr]
}%
\end{feynrules}